\documentclass[useAMS, usenatbib]{mn2e}
\usepackage{graphicx}
\usepackage{times}
\usepackage{color}
\newcommand{\aap}{A\&A}
\newcommand{\apj}{ApJ}
\newcommand{\apjl}{\apj}
\newcommand{\pasp}{PASP}
\newcommand{\aj}{AJ}
\newcommand{\mnras}{MNRAS}
\newcommand{\apjs}{ApJS}
\newcommand{\aapr}{A\&AR}
\newcommand{\pasj}{PASJ}
\newcommand{\kmps}{\mathrm{km~s^{-1}}}

\providecommand{\tabularnewline}{\\}
 \newenvironment{lyxlist}[1]
   {\begin{list}{}
     {\settowidth{\labelwidth}{#1}
      \setlength{\leftmargin}{\labelwidth}
      \addtolength{\leftmargin}{\labelsep}
      }}
   {\end{list}}

\begin{document}

\title[Time-series Paschen-$\beta$ spectroscopy of SU
  Aurigae]{Time-series Paschen-$\beta$ spectroscopy of SU Aurigae} 

\author[R. Kurosawa et\,al.]{Ryuichi Kurosawa\thanks{E-mail: rk@astro.ex.ac.uk}, Tim J. Harries and Neil H. Symington\\ School of Physics, University of Exeter, Stocker Road, Exeter EX4 4QL}

\date{Dates to be inserted}

\pagerange{\pageref{firstpage}--\pageref{lastpage}} \pubyear{2004}

\maketitle

\label{firstpage}

\begin{abstract}
  
  We present time-series echelle spectra of the Pa$\beta$ line of the
  T~Tauri star SU~Aur, observed over three consecutive nights.
  The line shows strong variability ($\sim10$~per~cent)
 \textcolor{black}{over the velocity range ($100\,\kmps,\,420\,\kmps$)
  in the red broad absorption component,
   and weaker variability ($\sim2$~per~cent) over the velocity
  range ($-200\,\kmps,\, 0\,\kmps$) in the
  blue wing}. The variability in the velocity range
  ($-200\,\kmps,\, 0\,\kmps$) is correlated with that in
  ($200\,\kmps,\,400\,\kmps$), and the variability in these velocity 
  ranges anti-correlates with that in ($0\,\kmps,\,100\,\kmps$). 
  The mean spectrum from the second night shows a suggestion of a
  blue-shifted absorption component at about $-150\,\kmps$, similar to
  that found in the H$\alpha$ and H$\beta$ lines.  
  We find the position of the subpeak in the red absorption component changes
  steadily with time, and its motion modulates on half the
  rotational period. We also find that the modulation of the line
  equivalent width is associated with a half and 
  a third of the rotational period, which is consistent with the
  surface Doppler images of SU~Aur. 
  Radiative transfer models of a rotationally modulated
  Pa$\beta$ line, produced in the shock-heated magnetospheric
  accretion flow,  are also presented. Models with a magnetic dipole
  offset reproduce the overall characteristics of the observed line
  variability, \textcolor{black}{including the line equivalent width and the motion of the
  subpeak in the red absorption trough}. 
   
\end{abstract}

\begin{keywords}
stars:formation -- stars: individual: SU Aur -- circumstellar matter -- stars: pre-main-sequence
\end{keywords}


\section{Introduction }

\label{sec:Introduction}

T~Tauri stars (TTS) are young ($<\sim3\times10^{6}\,\mathrm{yrs}$,
\citealt{appenzeller:1989}) low-mass stars, and are thought to be the
progenitors of solar-type stars. Classical T~Tauri stars (CTTS)
exhibit strong H$\alpha$ emission, and typically have spectral types
of F--K. Some of the most active CTTS show emission in higher order
Balmer lines and metal lines (e.g.~\mbox{Ca\,{\sc ii}}, H and K). They also exhibit
an excess amount of continuum flux in the ultraviolet (UV) and
infrared (IR).

Line profile studies of CTTS show evidence for both outflows and
inflows, as seen in the blue-shifted absorption features in H$\alpha$
profiles (e.g.~\citealt{herbig:1962}) and the inverse P~Cygni (IPC)
profiles (e.g.~\citealt{kenyon:1994}; \citealt{edwards:1994})
respectively. Typical mass-loss rates of CTTS are about
$10^{-9}\,\mathrm{M_{\odot}\, yr^{-1}}$ to $10^{-7}\,\mathrm{M_{\odot}\,
  yr^{-1}}$ (e.g.~\citealt{kuhi:1964}; \citealt{edwards:1987};
\citealt{hartigan:1995}), and mass-accretion rates range from 
$10^{-9}\,\mathrm{M_{\odot}\, yr^{-1}}$ to $10^{-7}\,\mathrm{M_{\odot}\,
  yr^{-1}}$ (e.g.~\citealt{kenyon:1987}; \citealt*{bertout:1988};
\citealt{gullbring:1998}).

In the currently favoured model of accretion on to CTTS, the
accretion discs are disrupted by the stellar magnetosphere which
channel the gas from the disc on to the stellar surface
(e.g.~\citealt{uchida:1985}; \citealt{koenigl:1991};
\citealt{cameron:1993}; \citealt{shu:1994}).  This picture of the
accretion flows is supported by the observation that CTTS have
relatively strong ($\sim10^{3}\mathrm{G}$) magnetic fields 
(e.g.~\citealt*{johns-krull:1999}; \citealt{guenther:1996};
\citealt{symington:2004b}) and by radiative transfer models which
reproduce the gross characteristics of the observed profiles for some
TTS (e.g.~ \citealt*{muzerolle:2001}). 

SU~Aurigae is a bright ($J\sim7$,
\citealt{chakraborty:2004}) young star with a G2 spectral type
\citep{herbig:1960}, and has been classified either as a CTTS (e.g.~
\citealt{giampapa:1993}; \citealt{bouvier:1993}) or early-type TTS
(e.g.~\citealt{herbst:1994}). It is also the prototype `SU Aur type'
 \citep{herbig:1988}. Based on modelling of the spectral
energy distribution (SED), SU~Aur is known to have an accretion disc
($\sim400\,\mathrm{au}$) along with an outer envelope (e.g.~
\citealt{akeson:2002}).  Near-infrared coronagraphic observations of
\citet{nakajima:1995} and \citet{grady:2001} have shown that the
object is associated with a reflection nebula and outflows. Most
recently, \citet{chakraborty:2004} have presented high resolution
(0.3 arcsec) $J$-, $H$- and $K$-band adaptive optics images displaying
 nebulosity with a disc-like structure around SU~Aur.

SU~Aur has been a subject of a number of variability studies, mainly
in the optical, e.g.~ \citet{giampapa:1993}, \citet{johns:1995},
\citet{petrov:1996}, \citet{oliveira:2000} and most recently
\citet{unruh:2004}. The H$\alpha$ line, which is most commonly used
for variability studies of TTS, is produced over a large circumstellar
volume, and shows evidence of outflow (winds) as well as
magnetospheric inflow.  \citet{johns:1995} studied the variability of
H$\alpha$ and H$\beta$ profiles of SU~Aur, and found that the
modulation of the wind and the accretion components in the profiles
are approximately $180^{\circ}$ out of phase. This led them to suggest
that the magnetic field axis is slightly tilted with respect to the
rotation axis of the star. 

\textcolor{black}{The topology of the magnetic field is still in
  dispute.  Despite of the overall success of the dipole field
  geometry used in explaining many observed features of CTTS,
  \citet{johns-krull:2002} re-examined the validity of this
  assumption.  They found that observations did not support the
  relationships between mass, radius, and rotation period that
  magnetospheric accretion theory predicts under the assumption of a
  dipolar field, but found better agreement when a non-dipole field
  topology was used. In addition, for SU~Aur, \citet{unruh:2004} found
  a lack of the circular polarisation across some strong photospheric
  lines; once again in conflict with the expectations for a
  large-scale dipole.  Further, \citet{oliveira:2000} found the
  time-lagged behaviour in the variability of some optical lines, and
  proposed that this can be caused by the presence of an azimuthal
  component in the magnetosphere.}

Near-IR profiles, such as Pa$\beta$ and Br$\gamma$ are likely to
become important probes of accretion in CTTS and their lower-mass
counterparts, since they appear to suffer from less contamination by
outflows than H$\alpha$ (e.g.~\citealt{folha:1998};
\citealt{folha:2001}, although see \citealt{whelan:2004}) and are
less affected by dust obscuration, enabling the study of accretion in
highly embedded objects. Here we present time-series Pa$\beta$ spectra
of SU~Aur, which we use to probe the accretion geometry. In
Section~\ref{sec:Observations} we give details on the observations
and the data reduction. The results and analysis are presented in
Section~\ref{sec:Results}. Radiative transfer models are computed in
an attempt to explain the overall observed variability feature in
Section~\ref{sec:Models}. The conclusions are given in
Section~\ref{sec:Conclusions}.


\section{Observations}

\label{sec:Observations}

A total of 503 Pa$\beta$~($\lambda=1.28181\,\mathrm{\mu m}$) spectra
with a signal-to-noise ratio (S/N) of $\sim90$ per pixel (in the
continuum) were obtained over 3 nights (2002 December 1--3) at the
United Kingdom Infrared Telescope (UKIRT) on Mauna Kea, Hawaii. The
approximate temporal sampling rate was one spectrum every 3
minutes.

The Cooled Grating Spectrometer No.~4 (CGS4), a 1--5~$\mathrm{\mu m}$
multi-purpose 2-D grating spectrometer with a $256\times256$ InSb
array, with the echelle grating and long camera ($300\,\mathrm{mm}$), 
was used. This provides a velocity resolution (at $1.28181\,\mathrm{\mu
  m}$) of $\sim16\,\mathrm{km\, s^{-1}}$ using $\sim1$~pix slit width
without the Nyquist stepping. The wavelength range of the echelle
spectra is about $3600\,\mathrm{km\, s^{-1}}$ providing enough
continuum on either side of the line for accurate rectification.
Unfortunately there are insufficient photospheric features to perform
a veiling correction \textcolor{black}{
(but see Section~\ref{sub:model-configuration}).
} 
Telluric standard stars were observed throughout each night (every 2--3 hours) to
correct for instrument response and atmospheric transmission.

The data obtained with CGS4 were reduced using a standard procedure
(c.f.~\citealt{puxley:1992}): 1.~bias subtraction, 2.~flat-field, and
3.~optimal extraction (\citealt{horne:1986}). All spectra were reduced
in the same manner, using the ORAC-DR pipeline program. Intrinsic
absorption components in the standard stars, used in step~3 to correct
for the telluric lines, were subtracted from the original spectra by
using the least square fit of the spectra with a combination of two
Gaussian functions and a third order polynomial. After the preliminary
reduction by the pipeline was finished, a wavelength calibration was
performed using a combination of OH sky emission lines and
comparison arc lines. Finally we applied heliocentric velocity
corrections and shifted the spectra into the rest-frame of SU~Aur,
adopting the radial velocity measurement ($+16\,\mathrm{km\, s^{-1}}$)
of \citet{herbig:1988}.


\section{Results}

\label{sec:Results}

\begin{figure}

\begin{center}

\includegraphics[%
  bb=0bp 0bp 489bp 704bp,
  scale=0.48]{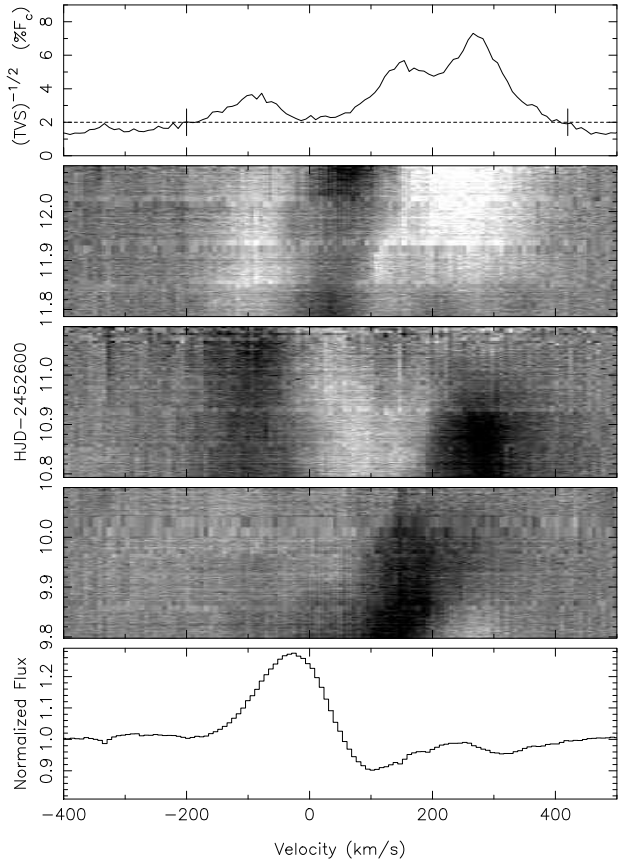}

\end{center}

\caption{The time-series spectra of Pa$\beta$. The mean  spectrum
  (normalised) of the 503 spectra obtained during the three night
  (2002 December 1--3) is shown at the bottom. In the middle panels,
  the quotient spectra (divided by the mean spectrum) as a function of
  time ($\mathrm{HJD-2452600}$) are plotted as greyscale images with the
  colour scaled from $1.1$ (white) to $0.9$ (black). The temporal
  deviation spectrum $(\mathrm{TVS})^{1/2}$ from the 503 spectra is
  shown on the top panel (c.f. Section~\ref{sub:Temporal-variance-spectra}). 
  In all plots, the horizontal axes are in
  velocity space in the stellar rest frame. The greyscale images
  clearly show variability on both red and blue sides of the
  spectra, but little variability is seen near the line centre. The
  dashed line in the top panel shows the statistical significance at the
  1 per cent level, and it indicates that the velocity range
  ($-200~\mathrm{km~s^{-1}}$, $420~\mathrm{km~s^{-1}}$) has 
  significant variability.}

\label{fig:threenights}

\end{figure}

A general overview of the dataset is presented in
Fig.~\ref{fig:threenights}. The mean spectrum shows a classic inverse
P~Cygni (IPC) profile, with the blue-wing of the emission component
extending to $-200$\,$\mathrm{km\, s^{-1}}$ and the absorption
component with maximum depth at $+100$\,$\mathrm{km\, s^{-1}}$ and
extending out to $+200$\,$\mathrm{km\, s^{-1}}$. The peak intensity of
the line, relative to the continuum, is about 1.25. Although
comparable in strength, the Pa$\beta$ profiles of SU~Aur presented by
\cite{folha:2001} show significant emission red-wards of the
absorption dip, and also have maximum absorption occurring at
$0$\,$\mathrm{km\, s^{-1}}$ i.e. they appear to be shifted by
$\sim90$\,$\mathrm{km\, s^{-1}}$ blueward compared to our dataset.

\begin{figure}

\begin{center}

\includegraphics[%
  clip,
  scale=0.35]{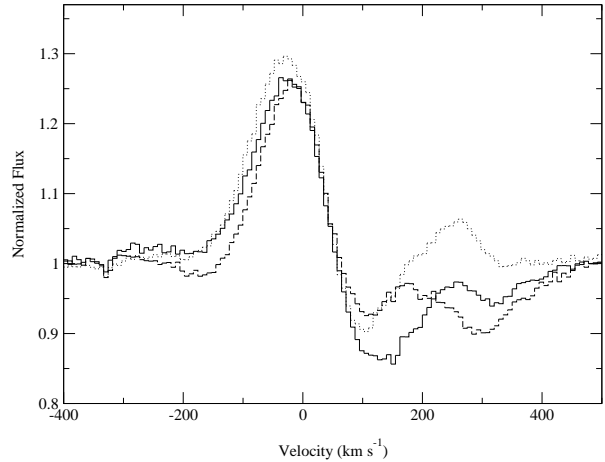}

\end{center}

\caption{Mean spectra from each night: The first night (solid),
  the second night (dashed), and the third night (dotted). While most
  of the variability occurs in the red side
  ($50~\mathrm{km~s^{-1}}$, $500~\mathrm{km~s^{-1}}$) of the line,
  small yet noticeable changes in flux level can be seen in
  ($-200~\mathrm{km~s^{-1}}, 0~\mathrm{km~s^{-1}}$). The flux level at
  $\sim250~\mathrm{km s^{-1}}$ is below the continuum on the first and the
  second night, but it is in emission (above the continuum) on the third
  night. A hint of a blue-shifted absorption component at
  $\sim 150\,\kmps$ is seen on the second night.} 

\label{fig:mean_sp}

\end{figure}

The mean spectra from each night are shown in Fig.~\ref{fig:mean_sp}.
It is immediately apparent that the blue side of the profile is more
stable than the red, with variability of less than 10 per cent (in
continuum units) at the line peak and blueward. There is a suggestion
of blue-shifted absorption at about $-150$\,$\mathrm{km\, s^{-1}}$ on
the second night.  Interestingly, the blue-shifted absorption
feature at $-150\,\kmps$ is also seen in H$\alpha$
\citep{giampapa:1993} and H$\beta$ \citep{johns:1995} profiles, and
the modulation period of the absorption feature is identified as the
rotational period of SU~Aur ($\sim3$~d). The red side of the mean
profiles shows strong variability, extending redward to over
$+400$\,$\mathrm{km\, s^{-1}}$. On the final night there is emission
on the red side of the absorption trough, and the profile more closely
resembles that presented by \cite{folha:2001}.

The greyscale dynamic spectra (Fig.~\ref{fig:threenights}) reveal the
gradual changes that occur over a timescale of hours. On the first
night there is a feature that accelerates monotonically
redward across the absorption trough. On the second night both the
red and blue sides of the profile are lower than the mean. Finally,
on the third night we see the growth of an emission feature in the
absorption dip. The characteristic level of this red-wing variability
is about 10 per cent. 

\subsection{Temporal variance spectra}

\label{sub:Temporal-variance-spectra}

In order to quantify the level and significance of the variability
seen in the Pa $\beta$ profiles, the temporal variance spectrum ($\mathrm{TVS}$)
analysis method proposed by \citet*{fullerton:1996} was applied to the
time series spectra from each night separately, and also to all three
nights combined. This method statistically compares the deviations in
line features with those of the adjacent continuum.  In other words,
the level of deviation at a given wavelength is computed with a weight
function which is inversely proportional to the signal-to-noise (S/N)
level of the continuum. By using the definition of
\citet{fullerton:1996}, the $j$-th velocity bin of the $\mathrm{TVS}$, in the
limit of high photon-count, can be written as:

\begin{equation}
\left(\mathrm{TVS}\right)_{j}
    =\frac{1}{N-1}\,\sum_{i=1}^{N}\left(S_{ij}-\bar{S_{j}}\right)^{2}\,\frac{w_{i}}{S_{ij}}\,
\label{eq:definition_TVS}
\end{equation}
where $S_{ij}$ is the signal of the $j$-th velocity bin belonging
to the $i$-th spectrum of a time-series, and $N$ is the number of observed
spectra.  $\bar{S_{j}}$ is the
weighted mean spectrum defined as 

\begin{equation}
\bar{S_{j}}=\frac{1}{N}\sum_{i=1}^{N}w_{i}S_{ij}\,\,.
\label{eq:weighted_mean_spectrum}
\end{equation}
The weighting factor $w_{i}$ for the $i$-th spectrum is defined as 

\begin{equation}
w_{i}=\frac{\left(1/\sigma_{ic}^{2}\right)}{\frac{1}{N}\sum_{k=1}^{N}\left(1/\sigma_{kc}^{2}\right)}
\label{eq:weight_noise}
\end{equation}
where $\sigma_{ic}$ and $\sigma_{kc}$ are the noise in the continuum
for the $i$-th and $k$-th spectra.

As mentioned by \citet{fullerton:1996}, a more convenient quantity to be
plotted is the square root of $\mathrm{TVS}$ (temporal deviation spectrum) because
$\left(\mathrm{TVS}\right)^{1/2}$ scales linearly with the spectral deviations;
hence, it provides a more accurate impression of the deviation from the
variability with respect to the continuum.

\begin{figure*}

\begin{center}

\begin{tabular}{ccc}
\includegraphics[%
  scale=0.32]{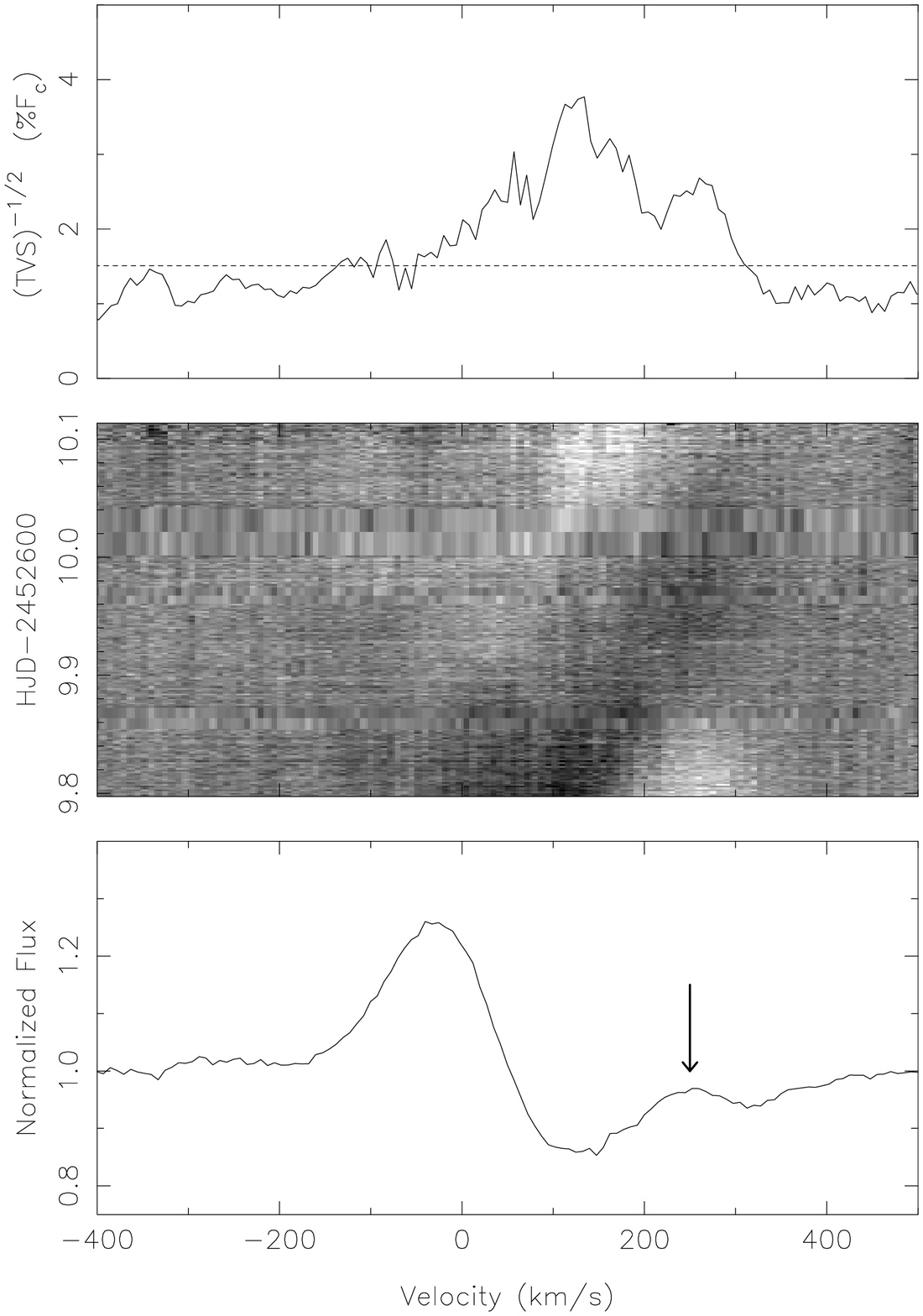}&
\includegraphics[%
  clip,
  scale=0.32]{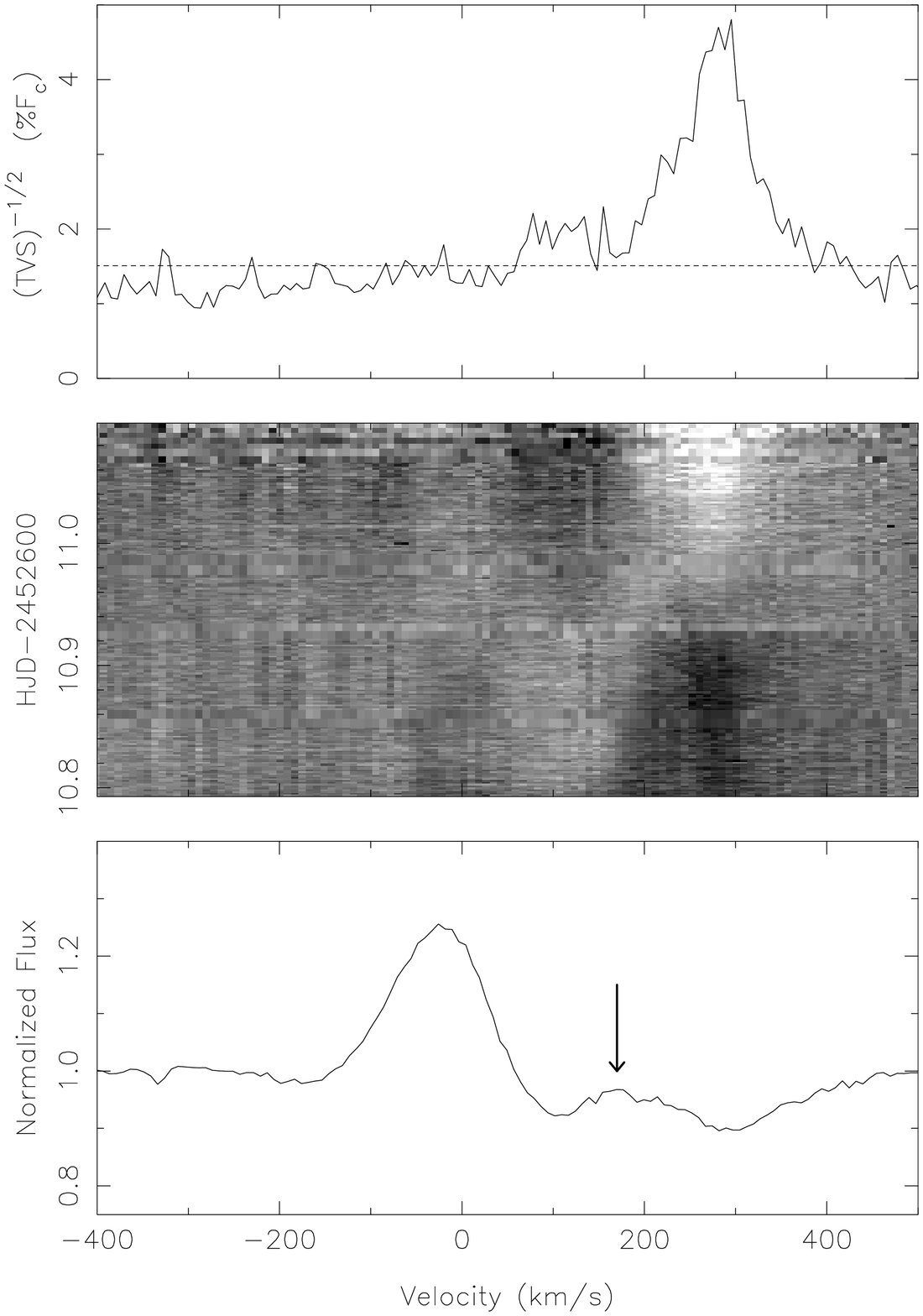}&
\includegraphics[%
  clip,
  scale=0.32]{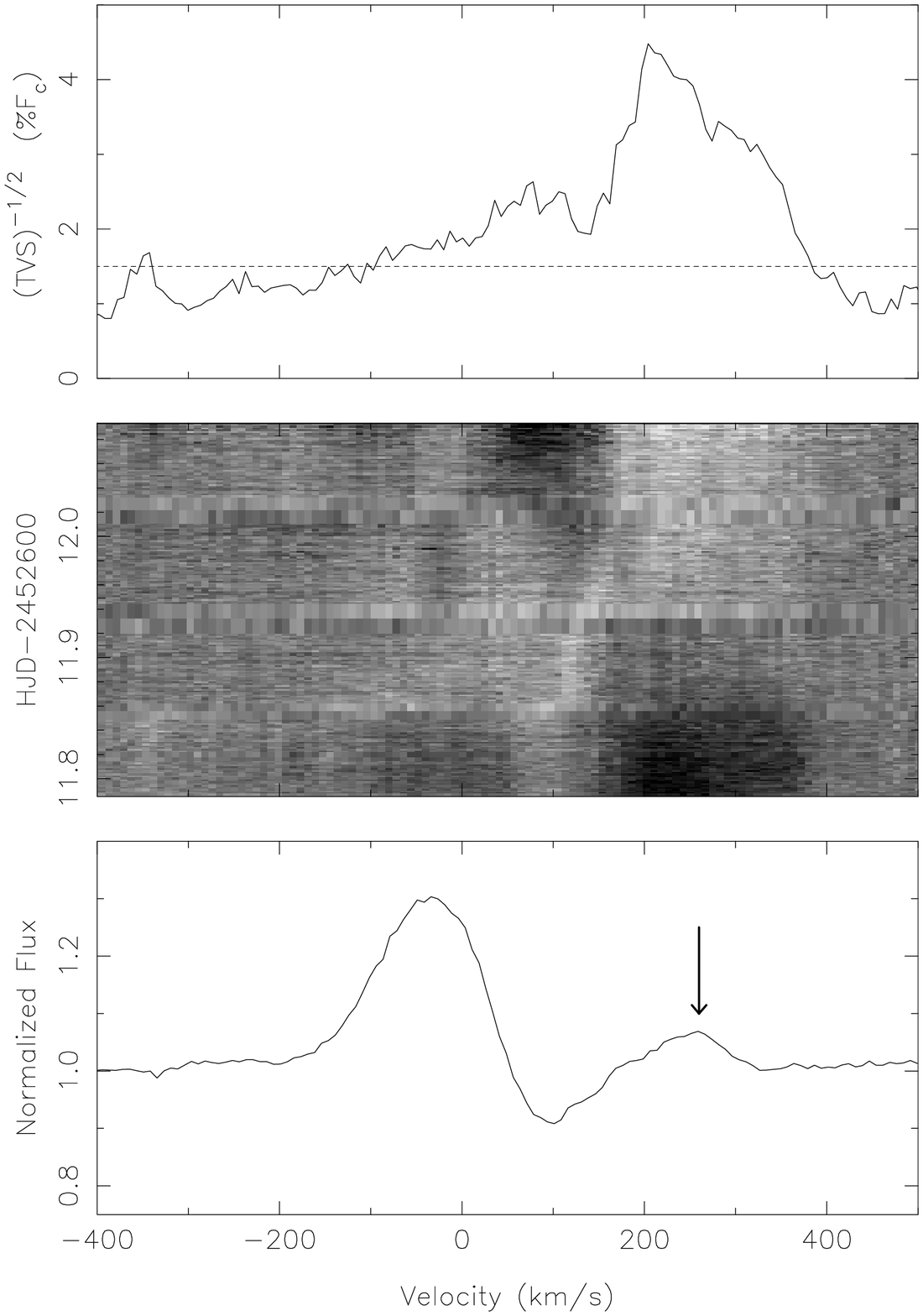}\tabularnewline
\end{tabular}

\end{center}

\caption{The summary of observations from each night (the first night
  to the third night from left to right respectively). For each night,
  the normalised mean spectrum (bottom), the quotient spectra (divided
  by the mean spectrum) (middle) with colour scaled from $1.08$
  (white) to $0.92$ (black) as a function of time
  ($\mathrm{HJD}-2452600$), and the temporal deviation spectrum
  $\mathrm{\mathrm{TVS}}^{-1/2}$  (top) in units of 1/100 of the continuum 
  flux ($\%F_{c}$) are shown. The dashed line in the top plot
  indicates the statistical significance for 1 per cent level. In the
  bottom plots, the positions of the subpeak in the red absorption
  troughs are indicated with arrows, and they seem to be moving from
  night to night (see Section~\ref{sub:motion-of-subpeak}).} 

\label{fig:eachnight}

\end{figure*}

Using equation~\ref{eq:definition_TVS}, the
temporal deviation spectrum of the times-series spectra from
each night was computed, 
and the results were placed in the top panels of
Fig.~\ref{fig:eachnight}.  In the same figure, the spectra divided by
the mean spectrum of each night (the quotient spectra) and the mean
spectrum are shown as a function of time. Similarly,
$\left(\mathrm{TVS}\right)^{1/2}$ computed from 
all three nights is placed in the top of Fig.~\ref{fig:threenights}.

The $\left(\mathrm{TVS}\right)^{1/2}$ value which corresponds to one per cent
statistical significance (the $\chi^{2}$-probability) is also shown in
the same figure as a reference point. In the absence of real
variability, the observed $\left(\mathrm{TVS}\right)^{1/2}$ values should be
below this level for 99 per cent of time. From the
$\left(\mathrm{TVS}\right)^{1/2}$ plot in Fig.~\ref{fig:threenights}, we
confirm that the line displays strong variability 
\textcolor{black}{within the velocity range ($100\,\mathrm{km\,
s^{-1}},\,420\,\mathrm{km\, s^{-1}}$) in the red wing, and weaker
variability within the velocity range ($-200\,\mathrm{km\,
  s^{-1}},\, 0\,\mathrm{km\, s^{-1}}$) in the blue wing.}
 The $\left(\mathrm{TVS}\right)^{1/2}$
plots in Fig.~\ref{fig:eachnight} show that the velocity ranges in
which the variability is most active change from night to night.

Using the mass ($M_{*}=2.25\,\mathrm{M_{\odot}}$) and the radius
($R_{*}=3.6\,\mathrm{R_{\odot}}$) of SU~Aur from \citet{cohen:1979},
we find its escape velocity to be
$V_{\mathrm{\mathrm{esc}}}=\left(2GM_{*}/R_{*}\right)^{1/2}=490\,\mathrm{km\,
  s^{-1}}$. This value is an upper-limit for the velocity of the
material as it impacts on the stellar surface, and indeed we see that
the extent of the red absorption in the line profile is
$\sim420\,\mathrm{km\, s^{-1}}$, and that the maximum velocity of
significant variability is $\sim420\,\mathrm{km\, s^{-1}}$.

\subsection{Motion of a subpeak in the absorption trough}

\label{sub:motion-of-subpeak}

\begin{figure}

\begin{center}

\includegraphics[%
  clip,
  scale=0.39]{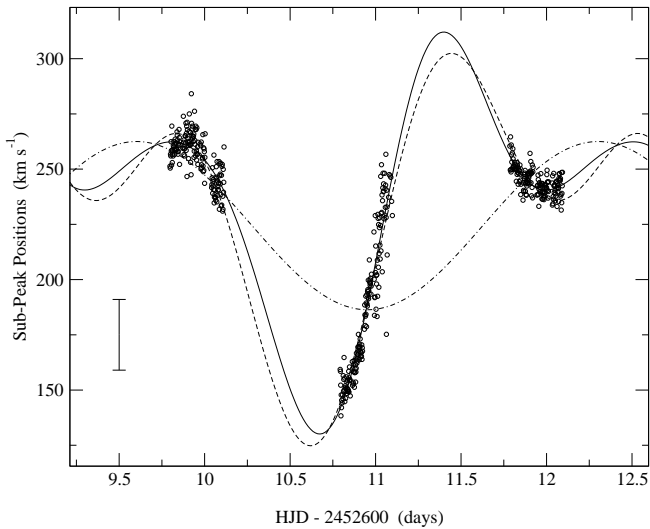}

\end{center}

\caption{The positions of the subpeak (circles) in the red
  wing of the Pa$\beta$ profile from the three nights are plotted as a
  function of time. The data were fitted with the Fourier series
  (equation~\ref{eq:fourier_series}) with the terms up to
  $n_{\mathrm{max}}=1$  (dash-dot line), $n_{\mathrm{max}}=2$ (dashed
  line) and $n_{\mathrm{max}}=3$ (solid line).  The corresponding
  fitting coefficients are summarised in
  Table~\ref{tab:subpeak_fit_parameters}. It is clearly shown that the
  line with $n_{\mathrm{max}}=1$ does not fit the data very well, but 
  the lines with $n_{\mathrm{max}}=2$ and $n_{\mathrm{max}}=3$ 
  fit the data equally well. The contribution from the third harmonic
  terms in the Fourier series is very small. 
  The motion of the subpeak in the red absorption trough is
  associated not only with the rotational period of SU~Aur ($P$),
  but also with half of the period ($P/2$). A typical size of the
  uncertainty in the peak positions is indicated on the lower-left corner.}  

\label{fig:motion-of-subpeak}

\end{figure}

\begin{figure}

\begin{center}

\includegraphics[%
  clip,
  scale=0.57]{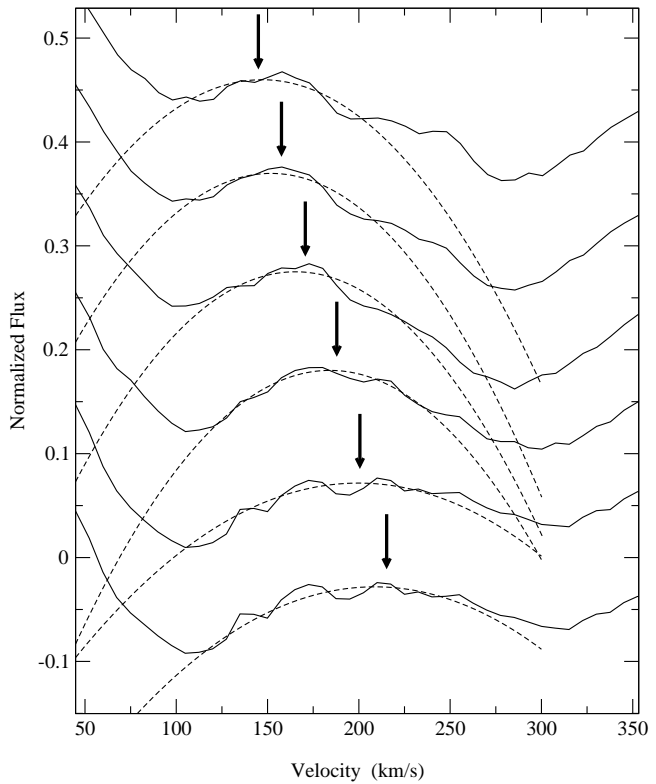}

\end{center}

\caption{\textcolor{black}{The time-series spectra of the Pa$\beta$
 from the second night (solid) and the quadratic fits for the subpeak positions
 (dashed). Each spectrum is the average of 27 consecutive spectra 
 ($\Delta t = 70$~min) in chronological order from the top to bottom.
  Each spectrum is vertically displaced by -0.1 from the previous
  spectrum for clarity. The position of a weak subpeak in the
  red-wing appears to be moving from $V \sim 170 \kmps$ to $\sim 220
  \kmps$ during the course of the second night. The positions of the
  subpeak were estimated using the original spectra (before being
  averaged over 27 spectra), by applying a quadratic fit to the data
  points around the subpeak (Fig.~\ref{fig:motion-of-subpeak}).  The
  arrows indicate the measured positions of the subpeak.}}  

\label{fig:subpeak-second-night}

\end{figure}

In the red absorption trough of the mean spectra (see
the bottom panels of Fig.~\ref{fig:eachnight}), we find the
presence of a subpeak at \textcolor{black}{$V=\sim250$}, $\sim180$ and $\sim260\,\kmps$
for the first, second and third nights.  We also find the position of
the subpeak (indicated by arrows in the same figure) changes during
one night. It appears that a main cause of the variability seen in the
red-wing (as seen in the greyscale images in the same figures) is due
to the motion of this subpeak, 

The positions of the subpeak were estimated for each spectrum, by applying
a quadratic fit to the data points around the subpeak. The results
are shown in Fig.~\ref{fig:motion-of-subpeak} as a function of time. 
The amplitude of the change in the subpeak position is
relatively large on the second night compared to those of the first
and the third nights.  
\textcolor{black}{
To demonstrate the motion of the subpeak occurred during the second
night, we plot the time-averaged ($\Delta t=70$~min) spectra in
Fig.~\ref{fig:subpeak-second-night}.  The figure shows a weak
subpeak moving from $V \sim 170 \kmps$ to $\sim 220 \kmps$ during the
course of the night.}

The subpeak positions were fitted with a function ($f$)
in the form of the Fourier series, 

\begin{equation}
f\left(t\right)=\frac{a_{0}}{2}+\sum_{n=1}^{n_{\mathrm{max}}}\left\{
a_{n}\cos\left(\frac{2\pi n}{P}t\right)+b_{n}\sin\left(\frac{2\pi
  n}{P}t\right)\right\} 
\label{eq:fourier_series}
\end{equation}
where $t$ and $P$ are the time of observation and the rotational
period of SU~Aur respectively, and $n_{\mathrm{max}}=3$.  
The reasons for choosing the
form of the fitting function above are as follows: 1.~a simplest
possible form is desirable, and 2.~the first harmonic terms in the Fourier
series might be present if the variation is related with the stellar
rotation, 3.~the second harmonic terms in the Fourier series might be
present if the variation is caused by the tilted-axis magnetospheric
accretion model (e.g.~see \citealt{shu:1994}; \citealt{johns:1995}),
and 4.~the third harmonic terms in the Fourier series might be present
because the surface Doppler images of SU~Aur by \citet{petrov:1996},
\citet{mennessier:1997} and \citet{unruh:2004} weakly suggest the
presence of cool spots on the surface at three different longitudes
which are approximately equally spaced.

The fit with $n_{\mathrm{max}}=3$ is shown in
Fig.~\ref{fig:motion-of-subpeak} along with the fits with
$n_{\mathrm{max}}=1$ and $n_{\mathrm{max}}=2$ for comparison. In all
cases, the fixed period of $2.7\,\mathrm{d}$ (\citealt{unruh:2004})
was adopted. See Table~\ref{tab:subpeak_fit_parameters}
for the summary of the fitting parameters. The figure clearly shows
that the fit with $n_{\mathrm{max}}=1$ fails to represent the data. 
On the other hand, the fits with $n_{\mathrm{max}}=2$ and
$n_{\mathrm{max}}=3$ are significantly better.

The two lines ($n_{\mathrm{max}}=2$ and $n_{\mathrm{max}}=3$) are very
similar to each other as we can see from the figure and the values of
the Fourier coefficients.  The contributions of the third harmonic
terms ($a_{3}$ and $b_{3}$) are very small compared to the first and
the second harmonic terms. 
To test whether the third harmonic terms should be included in the
fitting functions, we have performed an F-test
(c.f.\citealt{bevington:1969}). From the table, the numbers of degree
of freedom are $\nu_{2}=497$ and $\nu_{3}=495$ for 
$n_{\mathrm{max}}=2$ and $n_{\mathrm{max}}=3$ respectively. Using the
reduced chi-square values in the table, we find 
$F_{\chi}=\left\{\chi^{2}(n_{\mathrm{max}}=2)-\chi^{2}(n_{\mathrm{max}}=3)\right\}/\chi_{\nu_{2}}^{2}
= 2$ and 
$F=\chi_{\nu_{2}}^{2}/\chi_{\nu_{3}}^{2}=1$.  Since $F_{\chi}>F$, the third
harmonic terms should be included in the fitting although their
contribution is relatively small.

From this analysis we found that the motion of the
subpeak in the red absorption trough is possibly associated not only
with $P$, but also with $P/2$. This is consistent with the the
tilted-axis magnetospheric accretion model mentioned earlier. 

In spite of the excellent fits of the data with Fourier series, we
should not ignore a possibility that the rather large
amplitude of the change in the subpeak position seen on the second
night was caused by a single episodic event (e.g. by a temporal and
local enhancement of the accretion flow). Assuming 
conservation of energy,  the free-fall time ($t_{\mathrm{ff}}$) of an
object from the co-rotation radius ($r=R_{\mathrm{c}}$) of the star to
the stellar surface ($r=R_{*}$) can be written as
\begin{equation}
   t_{\mathrm{ff}}  =
    -\frac{1}{V_{\mathrm{esc}}
    \left(R_{\mathrm{c}}\right)}
       \int_{R_{\mathrm{c}}}^{R_{*}}\left(\frac{R_{\mathrm{c}}}{r}-1\right)^{-1/2}dr
   \label{eq:freefall_integral}
\end{equation}
 where $V_{\mathrm{\mathrm{esc}}}\left(R_{c}\right)=\left(2GM_{*}/R_{\mathrm{c}}\right)^{1/2}$,
  and $R_{*}$ and $M_{*}$ are the stellar radius and mass respectively.
 By evaluating the integral, one obtains 
\begin{equation}
   t_{\mathrm{ff}}  = 
     \frac{ R_{\mathrm{c}} }{ V_{\mathrm{esc}}\left(R_{\mathrm{c}}\right) }
     \left\{ \frac{\pi}{2} + q^{1/2}\left( 1-q \right)^{1/2} - \arcsin{\left(q^{1/2}\right)} \right\}
   \label{eq:freefall-time}
\end{equation}
where $q=R_{*}/R_{c}$.
Using the mass ($M_{*}=2.25\,\mathrm{M_{\odot}}$) and the radius
($R_{*}=3.6\,\mathrm{R_{\odot}}$) of SU~Aur from \citet{cohen:1979} 
along with the co-rotation radius ($R_\mathrm{c} = 3\,R_{*}$) in
equation~\ref{eq:freefall-time}, we find
$t_{\mathrm{ff}} = 10.5$~h.  This time-scale is comparable to the
observing time span of a given night, so the association of the
subpeak motion on the second night with a single episodic event cannot
be ruled out by timescale arguments alone. However, the systematic
trends in the acceleration of the feature on the first and third
nights strongly suggests that the feature is produced by rotational
modulation.  To exclude the possibility, the object must be observed
for a few rotational periods.  


\begin{table*}

\begin{minipage}{140mm}
\begin{center}

\caption{The summary of the fitting parameters in
  equation~\ref{eq:fourier_series} for the positions of the subpeak
shown in Fig.~\ref{fig:motion-of-subpeak}. $N$ and $\chi^{2}_{\nu}$
are the number of data points and the reduced chi-square respectively.}  

\label{tab:subpeak_fit_parameters}

\begin{tabular}{lrrrrrrrrr}
\hline 
$n_{\mathrm{max}}$&
$N$&
$\chi_{\nu}^{2}$&
$a_{0}$&
$a_{1}$&
$b_{1}$&
$a_{2}$&
$b_{2}$&
$a_{3}$&
$b_{3}$\tabularnewline
\hline
1&
$503$&
$3.5$&
$224.$&
$-35.7$&
$-13.2$&
$\ldots{}$&
$\ldots{}$&
$\ldots{}$&
$\ldots{}$\tabularnewline
$2$&
$503$&
$2.8$&
$288.$&
$-46.2$&
$36.6$&
$-36.7$&
$29.9$&
$\ldots{}$&
$\ldots{}$\tabularnewline
$3$&
$503$&
$2.8$&
$234.$&
$-42.3$&
$33.9$&
$-39.4$&
$28.2$&
$-9.14$&
$-1.06$\tabularnewline
\hline
\end{tabular}

\end{center}
\end{minipage}

\end{table*}

\subsection{Auto-correlation map}

\label{sub:Auto-correlation-map}

\begin{figure}

\begin{center}

\includegraphics[%
  bb=51bp 141bp 572bp 635bp,
  clip,
  scale=0.48]{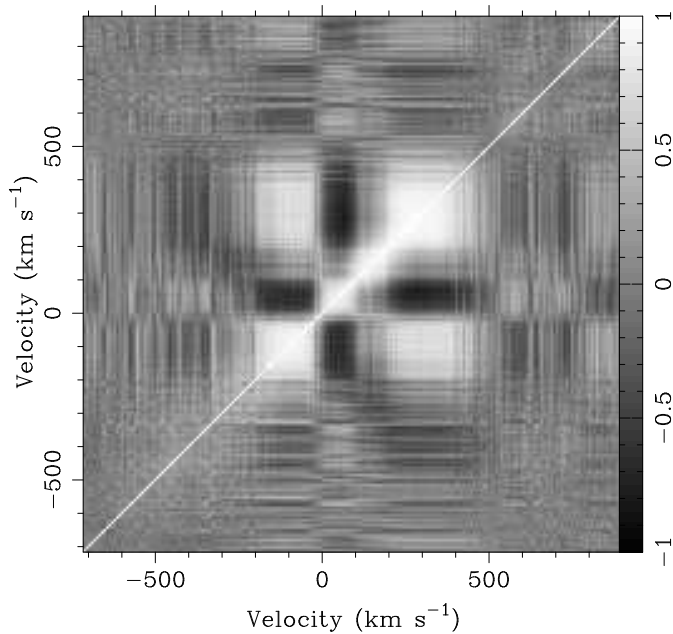}

\end{center}

\caption{Auto-correlation map of the Pa$\beta$ spectra (3 nights) with
  no time delay. The stronger the correlation, the brighter the pixels
  in the map. Similarly, the the stronger the anti-correlation, the
  darker the pixels are.  The flux changes seen between
  ($-200~\mathrm{km~s^{-1}}$, $0~\mathrm{km~s^{-1}}$)  and
  ($200~\mathrm{km~s^{-1}}$, $400~\mathrm{km~s^{-1}}$) are correlated
  with each other.  On the other hand,  the flux changes seen between
  those wavelength ranges weakly anti-correlate with the flux changes
  seen in ($0~\mathrm{km~s^{-1}}$, $100~\mathrm{km~s^{-1}}$).} 

\label{fig:auto-correlation}

\end{figure}

The auto-correlation map for the Pa$\beta$ spectra
(with all 503 spectra) was calculated to examine if variability at a given velocity
bin is correlated with those at a different part of spectra. The result
was placed in Fig.~\ref{fig:auto-correlation} as a greyscale
image. The correlation coefficient value ($C_{ij}$) of the $i$-th row
and the $j$-th column in the map was computed by using: 

\begin{equation}
    C_{ij} = \frac{\sum_{m=1}^{N}\left(S_{mi}-\bar{S_{i}}\right)\left(S_{mj}-\bar{S_{j}}\right)}
                         {\sum_{m=1}^{N}\left(S_{mi}-\bar{S_{i}}\right)^{2}} 
\label{eq:auto-correlation}
\end{equation}
where $S_{mn}$ is the signal of the $n$-th velocity bin in the $m$-th
spectrum of the time-series, and $N$ is the total number of
spectra. $\bar{S_{n}}$ is the mean signal of the $n$-th velocity bin
in the time-series.  The map is symmetric about the diagonal ($i=j$
pixels).  The range of the correlation values varies from $-1$ to $1$
which correspond to a strong anti-correlation and a strong correlation
respectively.

The map is useful to visually identify velocity bin regions that correlate
or anti-correlate with each other
(e.g. \citealt{johns:1995}; \citealt{oliveira:2000}).  
The figure shows the profile variability
in the velocity range ($-200\,\mathrm{km\, s^{-1}},0\,\mathrm{km\, s^{-1}}$)
weakly correlates with that for ($200\,\mathrm{km\, s^{-1}},400\,\mathrm{km\, s^{-1}}$).
On the other hand, they seem to weakly anti-correlate with the variability
in ($0\,\mathrm{km\, s^{-1}},100\,\mathrm{km\, s^{-1}}$). The pattern
seen in the map is very similar to that seen in the H$\beta$ auto-correlation
function of \citet{oliveira:2000}. They also found three similar
wavelength ranges in which the flux levels correlate and anti-correlate
with one another. 

The auto-correlation maps with time lags of $P/2$, $P/3$
and $P/4$ where $P$ is the rotational period were also computed
(with $P=2.7\,\mathrm{d}$ from \citealt{unruh:2004}). Because of
the poor time coverages (3 nights at a single location) of the data,
we were unable to draw any significant conclusion from those maps, and
they are not shown here.

\subsection{Line equivalent width}

\label{sub:Line-equivalent-width}
                                   
\begin{figure}

\begin{center}

\includegraphics[%
  clip,
  scale=0.49]{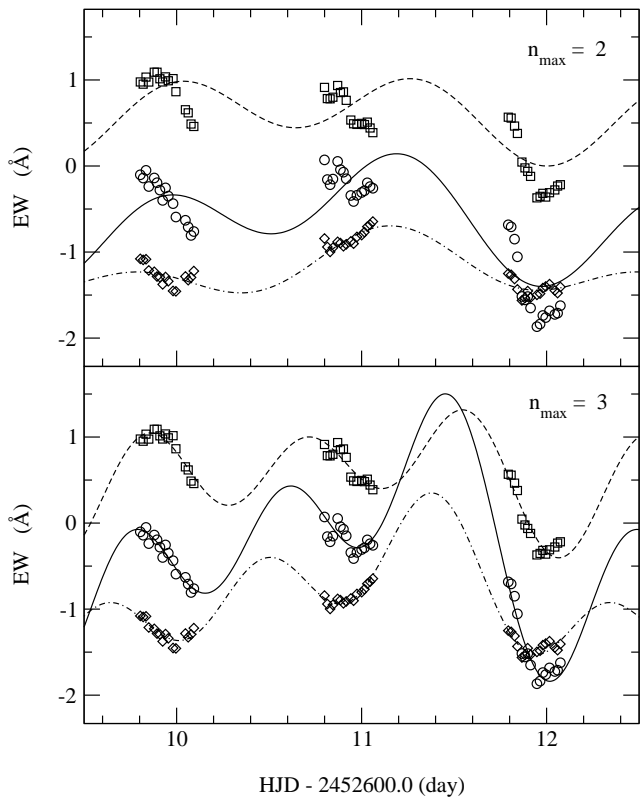}

\end{center}

\caption{\textcolor{black}{The equivalent
  widths (EW) of Pa$\beta$ using the velocity bins $-500~\kmps < V < 500~\kmps$
  (circles), $-500~\kmps <V< 0~\kmps$ (diamonds) and $0~\kmps < V < 500~\kmps$ (squares) are
  plotted as a function time.  The top panel shows the data fits using
  equation~\ref{eq:fourier_series} with $n_{\mathrm{max}}=2$, 
  and the bottom panel shows those with $n_{\mathrm{max}}=3$. The fits
  for the total EW ($-500~\kmps < V < 500~\kmps$), the blue EW
  ($-500~\kmps <V< 0~\kmps$) and the red EW ($0~\kmps < V < 500~\kmps$)
  are shown in solid, dashed and dash-dot lines respectively.
  The rotational period, $2.7~\mathrm{d}$ \citep{unruh:2004}, was kept constant in the
  fitting procedure.  Table~\ref{tab:fit_parameters} summarises the
  fitting parameters.}
}  

\label{fig:EW_flux_var}

\end{figure}

The line equivalent widths (EWs) of Pa$\beta$ profiles (shown in
Fig.~\ref{fig:threenights}) are computed by averaging/rebinning over
10 profiles obtained consecutively (approximately over 30 minutes), in
order to decrease the size of variance due to the relatively low
average S/N ($\sim90$) in the original profiles.  
\textcolor{black}{
Three different ranges of velocity
bins were used for computing the EWs.  The total EWs were computed using
the velocity range $-500~\kmps < V < 500~\kmps$.  The EWs of
the red wing (the red EWs) were computed using $0~\kmps < V <
500~\kmps$, and those of the blue wing (the blue EWs) with $-500~\kmps
< V < 0~\kmps$.}  The results are shown
in Fig.~\ref{fig:EW_flux_var}.  During the first night, the
\textcolor{black}{total} EW changes
between $-0.1$ and $-0.8$\,\AA. It changes between $+0.1$
and $-0.4$\,\AA\, on the second night, and during the third night it
changes between $-0.7$ and $-1.9$\,\AA. The negative sign in the
EW values indicates that the line is in emission. From night to night, the average
EW changes from $-0.4$ to $-0.2$\,\AA, and then to $-1.5$\,\AA. 

\textcolor{black}{
As we can see from this figure and also from the temporal
variance spectra in Fig.~\ref{fig:threenights}, the amplitude of the
variability for the red wing (or the red EW) is much larger than that
for the blue wing; hence, the basic behaviour of the total EW
variability follows that of the red wing (the red EW).
}

\textcolor{black}{
According to the figure, the variability of the total EW and that of
the red EW appear to be correlated with each other.  On other hand, the
variability of the total EW and that of the blue wing appear to be
anti-correlated for the first and the second night, but not on the
third night.  With much larger time steps ($\sim 1$d),
\citet{johns:1995} found a similar anti-correlation 
behaviour of the H$\alpha$ EWs measured using the velocity bins
$V\sim -150\,\kmps$ and those measured using $V\sim100\,\kmps$. This is
consistent with the auto-correlation map
(Fig.~\ref{fig:auto-correlation}) shown earlier.}

The data points were fitted with the Fourier series 
(equation~\ref{eq:fourier_series}) for $n_{\mathrm{max}}=2$ and
$n_{\mathrm{max}}=3$ cases separately.  As done for the fitting of the
the subpeak positions in Section~\ref{sub:motion-of-subpeak}, the
period was constrained to be $2.7\,\mathrm{d}$ 
(\citealt{unruh:2004}) in the fitting procedure.
The results are also shown in Fig.~\ref{fig:EW_flux_var}, and the
corresponding Fourier coefficients are summarised in
Table~\ref{tab:fit_parameters} along with the reduced chi-square
($\chi_{\nu}^{2}$) values. 

\textcolor{black}{
For all cases (total, red and blue EWs), 
the lines with $n_{\mathrm{max}}=3$ are better representations
of the data points, as we can see from the figure and the $\chi_{\nu}^{2}$ values in the
table. 
}
To test the validity of the statement above, we
performed an F-test. \textcolor{black}{For the total EWs fits,} 
the numbers of degrees of freedom are $\nu_{2}=42$ and $\nu_{3}=40$ for
$n_{\mathrm{max}}=2$ and $n_{\mathrm{max}}=3$
\textcolor{black}{(Table~\ref{tab:fit_parameters})} respectively. Using the 
values of the reduced chi-square values in the table, we find
$F_{\chi}=\left\{\chi^{2}(n_{\mathrm{max}}=2)-\chi^{2}(n_{\mathrm{max}}=3)\right\}/\chi_{\nu_{2}}^{2}
= 37$ and
$F=\chi_{\nu_{2}}^{2}/\chi_{\nu_{3}}^{2}=8.1$; hence,
$F_{\chi}>F$. From this test, we \textcolor{black}{found} that the inclusion of the
third harmonic terms are a real improvement in the fitting and they
should be included.  
\textcolor{black}{
Similar conclusions were found for the blue and
the red EW variability curves.
}

Interestingly, the contribution of the third harmonic terms are
as important as the first and the second harmonic terms as one can see
from the values of the Fourier coefficients in
Table~\ref{tab:fit_parameters}. This is consistent with the 
idea that variation is caused by the combination of the tilted-axis
magnetospheric accretion model (for the second harmonic terms) and the
presence of the cool spots of the surface at three different
longitudes which are approximately equally spaced from the surface
Doppler images (for the third harmonic terms), as briefly mentioned in
Section~\ref{sub:motion-of-subpeak}.  

\textcolor{black}{
Since the data sampling span of one night is very similar to $P/3$, it
is very difficult to distinguish the $P/3$ component found in the
Fourier analysis from a spurious detection.  To overcome this
shortcoming, a similar observation must be performed at multiple sites
(for continuous phase coverage) for at least a few rotational periods.
Also, authors would like to remind readers that the surface Doppler
images of SU~Aur obtained by \citet{petrov:1996} and 
\citet{mennessier:1997} suffer from the same data sampling problem as ours. 
The surface Doppler images constructed by \citet{unruh:2004}, on the other
hand, are based on the multi-site observations which has a fairly
continuous phase coverage and for a few periods; however, they has
significant problems with non-periodic line variability.
}

\begin{table*}

\begin{minipage}{140mm}
\begin{center}

\caption{
\textcolor{black}{The summary of the fitting parameters in
  equation~\ref{eq:fourier_series} for the equivalent width (EW) data
  shown in Fig.~\ref{fig:EW_flux_var}.  $N$ and  $\chi^{2}_{\nu}$ 
  are the number of data points and the reduced chi-square
  respectively. The velocity bins used for the total, red, and blue EWs
  are $-500~\kmps < V < 500~\kmps$, $0~\kmps < V < 500~\kmps$, and
  $-500~\kmps <V< 0~\kmps$ respectively.
}
} 

\label{tab:fit_parameters}
\begin{tabular}{rcrrrrrrrrrr}
\hline 
{}& $n_{\mathrm{max}}$& $N$& $\chi_{\nu}^{2}$& $a_{0}$& $a_{1}$& $b_{1}$& $a_{2}$& $b_{2}$& $a_{3}$& $b_{3}$\tabularnewline
\hline
Total& $2$& $48$& $3.1$& $-1.20$& $ 0.39$&  $-0.06$& $-0.24$& $ 0.41$& $\ldots{}$& $\ldots{}$ \tabularnewline
Total& $3$& $48$& $0.4$& $-0.36$& $ 0.59$&  $ 0.43$& $-0.52$& $ 0.20$& $0.18$&$-0.72$         \tabularnewline
Red&   $2$& $48$& $7.2$& $-1.21$& $-0.21$&  $ 0.06$& $ 0.29$& $-0.25$& $\ldots{}$& $\ldots{}$ \tabularnewline
Red&   $3$& $48$& $0.8$& $-1.17$& $-0.22$&  $-0.09$& $ 0.26$& $-0.11$& $-0.39$& $0.31$        \tabularnewline
Blue&  $2$& $48$& $0.7$& $ 2.38$& $-0.17$&  $-0.20$& $ 0.01$& $-0.23$& $\ldots{}$& $\ldots{}$ \tabularnewline
Blue&  $3$& $48$& $0.2$& $ 1.55$& $-0.37$&  $-0.34$& $ 0.25$& $-0.09$& $0.21$& $ 0.39$        \tabularnewline
\hline
\end{tabular}

\end{center}
\end{minipage}

\end{table*}

\section{Models}

\label{sec:Models}

These data, along with other spectroscopic time-series observations,
provide some strong constraints on the possible magnetospheric
geometry of SU~Aur. Specifically, in
Section~\ref{sub:motion-of-subpeak}, we have shown that the motion of
the subpeak in the red absorption trough is related not only to the
rotational period of the star, but also to half of the rotational
period.  This phenomenon is consistent with the tilted-magnetic axis
models (e.g.~see \citealt{shu:1994};
\citealt{johns:1995}) as mentioned earlier. Furthermore,
the analysis in Section~\ref{sub:Line-equivalent-width} suggests
that the variability of the Pa$\beta$ equivalent width is associated with a
 half and and a third of the rotational period of the star. This
characteristic is also recovered in  surface Doppler images by
\citet{petrov:1996}, \citet{mennessier:1997} 
and \citet{unruh:2004}, and is indicative of the presence of 
cool spots on the surface at three different longitudes which are
approximately equally separated.

To date the observational phenomena associated with SU~Aur have been
interpreted using cartoon-like models of the circumstellar geometry
and dynamics. Although useful, it is not clear that these necessarily
simplistic descriptions provide a reasonable explanation of the
changes in line-profile shape that are observed. However, it is also
true that the near-star geometry is sufficiently complicated that a
detailed fit to the observations is not currently tractable, and will
probably require a combination of simultaneous time-series
observations (spectroscopy, circular polarimetry and photometry) spanning a wide
range of wavelength. In this section we adopt a modelling approach
that, while falling short of a formal fit, provides a greater
quantitative insight into the line profile variability than a cartoon
description. Our intention is to develop radiative-transfer models
based on the simple magnetospheric geometries that have been proposed
in previous studies (\citealt{johns:1995}; \citealt{petrov:1996}). We
will be able to determine whether or not these geometries are capable
of reproducing the gross characteristics of the line profile
variability, and therefore make a better assessment of the
applicability of the models.

\subsection{The modelling code}

The three-dimensional Monte Carlo radiative transfer
code \textsc{TORUS} (\citealt{harries:2000}; \citealt{kurosawa:2004a};
\citealt*{symington:2004a}) is used to compute the Pa$\beta$ 
line profiles as a function of time (rotational phase). First, the
model computes the non-LTE populations of 14-level Hydrogen atoms in the 
magnetosphere which is funnelling the gas through the magnetic field
lines from the inner edge of the accretion disc.   Second, the model
computes the observed profile of Pa$\beta$ as a function of rotational
phase.  In our models, the Sobolev approximation
(c.f. \citealt{mihalas:1978}) is used in the framework of
core-plus-halo Monte Carlo radiative transfer method, in which the
photosphere is treated separately from the outer atmosphere
(e.g. accretion streams) of a star.  
Readers are referred to \citet{symington:2004a} for a detailed description
of the spectroscopic model of hydrogen lines arising from the
accretion streams in the magnetosphere. All of the accretion flow is
assumed to be constant in our models i.e.\, there is no periodic
pulsation of density enhancement etc. 

No rotational velocity
component is included in our models as it was neglected by
\citet{hartmann:1994} and \citet{symington:2004a}.
The effect of the
rotation on Pa$\beta$ is expected to be small according to the
calculations of \citet[][see their fig.~8]{muzerolle:2001}. 
\textcolor{black}{
A slightly larger amount of rotational broadening than that of \citet{muzerolle:2001}
is expected for SU~Aur since $v\sin{i}\sim60\,\kmps$
(\citealt{johns-krull:1996}; \citealt{unruh:2004}) while \citet{muzerolle:2001}
used $v\sin{i}\approx10\,\kmps$ in their calculation. 
Interestingly, \citet{oliveira:2000} found the time-lagged
behaviour in the variability of some optical lines, and proposed that
this can be caused by the presence of the azimuthal component in the
magnetosphere which might be caused by the interaction of the
rotating magnetosphere with the circumstellar disc. Although the
effect of the rotation on the Pa$\beta$ line may be important for a
quantitative measurement, it should not affect the conclusion drawn from
the qualitative analysis presented in this section. The exact effect
of the rotation on the line formation and variability will be
investigated more carefully in a future paper.} 

\subsection{Model configurations}

\label{sub:model-configuration}

A star with radius $R_{*}$ and mass $M_{*}$ is placed at
the origin ($O$) of a Cartesian coordinate system ($x$,$y$,$z$) as
shown in Fig.~\ref{fig:model-configuration}. The rotational axis of
the star is set to be identical to the $z$-axis, and the sense of
rotation is counter-clockwise when the star is viewed pole-on.  The
y-axis is perpendicular and into the page. An observer is placed on the $z$-$x$
plane with an inclination angle $i$ measured from the $z$-axis. The
structure of the magnetosphere is the same as that of 
\citet{hartmann:1994} i.e. the shape of the dipole (poloidal) magnetic
field lines is described by 

\begin{equation}
r=R_{\mathrm{m}}\,\sin^{2}\theta
\label{eq:dipole_field}
\end{equation}
where $r$, $R_{\mathrm{m}}$ and $\theta$ are the radial distance from the
centre of the star to a point ($p$) along a field line, the distance to the
point where the field line intersects with the equatorial plane, and
the polar angle measured from the magnetic axis 
(normally $z$-axis) respectively.  The magnetic axis can be inclined
by a small angle $\theta_{\mathrm{m}}$ with respect the rotational axis. 
We consider the following three models:

\begin{figure}
\begin{center}

\includegraphics[%
  clip,
  scale=0.45]{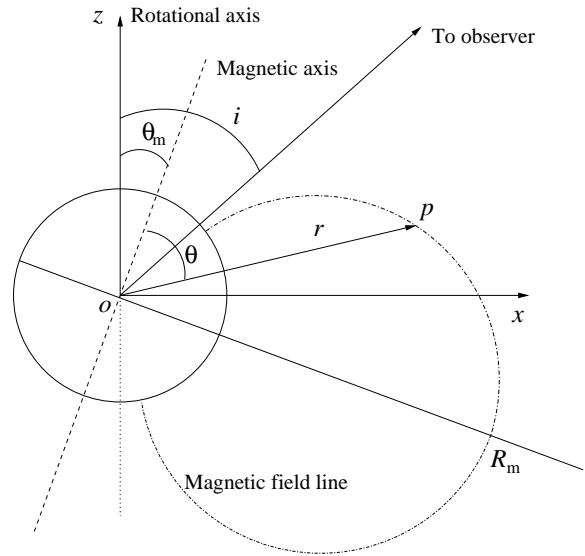}

\end{center}

\caption{Basic model configuration. A star radius $R_{*}$ and mass
$M_{*}$  is placed at the origin ($O$) of a Cartesian coordinate
system ($x$,$y$,$z$).  The rotational axis of the star is identical to
the z-axis, and the sense of rotation is counter-clockwise when the
star is viewed pole-on (viewed from $+z$ direction).  The $y$-axis is
perpendicular and into the page. An observer is placed on the $z$-$x$
plane with an inclination angle $i$ measured from $z$-axis. The
magnetic axis is tilted by a small angle $\theta{\mathrm{m}}$ with
respect to the rotational axis.  The shape of the dipole (poloidal)
magnetic field lines is described by
$r=R_{\mathrm{m}}\,\sin^{2}\theta$ as in
e.g. \citet{hartmann:1994}.}

\label{fig:model-configuration}

\end{figure}

\begin{figure}
\begin{center}

\includegraphics[%
  clip,
  scale=0.45]{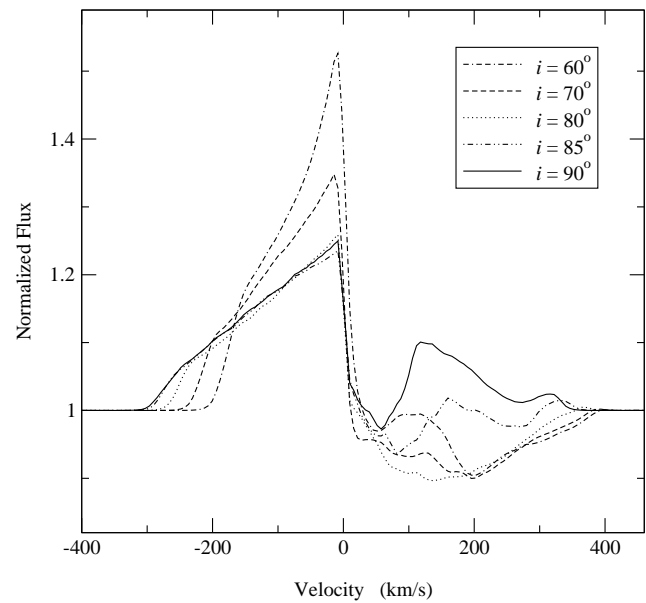}

\end{center}

\caption{The Pa$\beta$ line profiles computed at five different
inclinations ($i = 90^{\circ},\, 85^{\circ},\, 70^{\circ},\,
60^{\circ}$) using the same magnetosphere as in Model~A,
but without tilting the magnetic axis i.e. $\theta_{\mathrm{m}}=0$.  
The subpeak in the red wing appears in emission only at the high
inclination ($i>\sim 85^{\circ}$). At 
lower inclinations ($i<\sim 70^{\circ}$), the flux level of the blue
side of the profile is very sensitive to change in inclination. On
the other hand, at higher inclination angles  ($i>\sim 70^{\circ}$),
they are insensitive to the change in the inclination angle. }   

\label{fig:inclination-effect}

\end{figure}

\begin{lyxlist}{00.00.0000}
  \item [Model~A] A star is 
                  surrounded by the axi-symmetric magnetic field (described by
                  equation~\ref{eq:dipole_field}) with constant accretion flow along the
                  magnetic field lines, but the  magnetic axis tilted by
                  $10^{\circ}$ with respect to the rotational axis.
  \item [Model~B] A star is surrounded by the
                  axi-symmetric magnetic field (described by equation~\ref{eq:dipole_field})  
                  except three thin ($10^{\circ}$) gaps, where there is no magnetic field and
                  gas,  are longitudinally placed with equal separations
                 (i.e. $120^{\circ}$). The magnetic axis is aligned with the rotational axis. 
  \item [Model~C] A combination of Model~A and Model~B. The
                  magnetosphere has three $10^{\circ}$ gaps, and the
                  magnetic axis is tilted by $10^{\circ}$ with respect to
                  the rotational axis.
\end{lyxlist}

In all three models, we have adopted the following stellar
parameters:  the mass  $M_{*}=2.25\,\mathrm{M_{\odot}}$ and the radius
$R_{*}=3.6\,\mathrm{R_{\odot}}$  \citep{cohen:1979}. The
magnetospheric radius ($R_{\mathrm{m}}$) is assumed to be comparable
to the size of the co-rotation radius ($R_{\mathrm{c}}$) of the star
(e.g. \citealt{pringle:1972}; \citealt{ghosh:1979}; 
\citealt{shu:1994b};
\citealt{romanova:2002}). $R_{\mathrm{c}}\approx3.0~R_{*}$ for $R_{*}$
and $M_{*}$    values given above. The range of the magnetospheric
radius is chosen to be $R_{\mathrm{m}}=2.5-3.5\,R_{*}$, which is in
proportion to the small/narrow model of \citet{muzerolle:2001}. The
temperature structure along the magnetic field was adopted from
\citet{muzerolle:2001} with the maximum temperature
($T_{\mathrm{max}}$) of 8000~K (see their fig.~2).  The stellar
continuum of the core star is described by a model atmosphere of
\citet{kurucz:1979} with $T_{\mathrm{eff}} = 5750\,\mathrm{K}$ and
$\log{g}=4$ (cgs).

A simple geometrically thin and optically thick accretion disc is
placed just outside of the outer edge of the magnetosphere. All
the photons which encounter the disc are absorbed. 
\textcolor{black}{
  With this assumption of the disc, a photon emitted from the bottom half of
  the magnetosphere can be occulted by the disc, but a photon emitted
  from the top half can not be occulted by the disc. 
  This is reasonable for computing Pa$\beta$ profiles except for cases
  with very high inclination angles.  If a more realistic disc (e.g. a
  flared disc) was used in the model, we would expect the reduction of
  the photospheric continuum (e.g. \citealt{chiang:1999}) as well as
  the line emission due to obscuration by the outer part of the
  disc. Predicting the exact effects on the line profile shapes caused
  by this type of obscuration requires modelling of a
  self-consistent accretion disc, which is beyond the scope of this
  paper.
}

\textcolor{black}{Since we do not have the estimate of the amount of the
veiling around Pa$\beta$ line, the flux contribution from the disc itself is not
considered in our calculations.  A recent measurement of
the veiling at $2.2\mathrm{\mu m}$ for SU~Aur is $0.6\pm0.3$
\citep{muzerolle:2003}, but the value should be significantly smaller
at $1.3\mathrm{\mu m}$  (Pa$\beta$). If the veiling 
correction was taken into account, the line strength of the
models presented in this paper would be weaker.}
The mass-accretion rates ($\dot{M}$) for the models were
chosen so that the mean observed line strength is approximately
reproduced.

\textcolor{black}{The inclination is set to $i=80^{\circ}$ which is 
higher than the previously assumed values
(e.g. \citealt{muzerolle:2003}; \citealt{unruh:2004}).} 
Fig.~\ref{fig:inclination-effect} shows the
Pa$\beta$ line profiles computed at different inclination angles using
the same magnetospheric accretion structure as in Model~A, but with no dipole offset
i.e. $\theta_{\mathrm{m}}=0$.    At lower inclinations ($i<\sim
70^{\circ}$), the flux level of the blue side of the profile is very
sensitive to changes in inclination. This will cause too much
variability in the blue wing if $\theta_{\mathrm{m}}=10^{\circ}$ (as
in Model~A); hence it will be inconsistent with the observation
(c.f. Figs.~\ref{fig:threenights} and \ref{fig:eachnight}). On the
other hand, at higher inclination angles  ($i>\sim 70^{\circ}$), they
are insensitive to the change in the inclination angle. Interestingly,
we found that the higher inclination angle ($i>\sim 
85^{\circ}$ in Fig.~\ref{fig:inclination-effect}) is needed to have the
subpeak in the red absorption trough in emission 
(above continuum) as seen in the mean spectra of Pa$\beta$ from the
third night of the observation (Fig.~\ref{fig:eachnight}).

In Models~B and C, the locations of the gaps are at the azimuth angles
$105^{\circ}$, $225^{\circ}$ and $345^{\circ}$ (measured from
$+x$-axis before tilting) at time $t=0$ or equivalently $\mathrm{phase}=0$. 
For the models with the tilted magnetic axis (Models~A and C), the 
inclination of the magnetic axis with respect to the observer changes
from $90^{\circ}$ to $70^{\circ}$, then to $90^{\circ}$ as the
rotational phase changes from $0$ to $0.5$, then to $1.0$. 
The model parameters are summarised in Table~\ref{tab:model-parameters}.

\begin{table}

\caption{Model parameters} 

\label{tab:model-parameters}

\begin{tabular}{crcc}
\hline 
Model&
$\theta_{\mathrm{m}}$&
\# of gaps &
$\dot{M}$\tabularnewline
&
$\left(\,^{\circ}\right)$&
&
$\left(10^{-7}\,\mathrm{M_{\odot}}\,\mathrm{yr^{-1}}\right)$\tabularnewline
\hline
A&
$10$&
$0$&
$1.15$\tabularnewline
B&
$0$&
$3$&
$1.25$\tabularnewline
C&
$10$&
$3$&
$1.15$\tabularnewline
\hline
\end{tabular}

\end{table}

\subsection{Model spectra}
\label{sub:model-spectra}

\begin{figure*}

\begin{center}

\begin{tabular}{ccc}
\includegraphics[%
  scale=0.32]{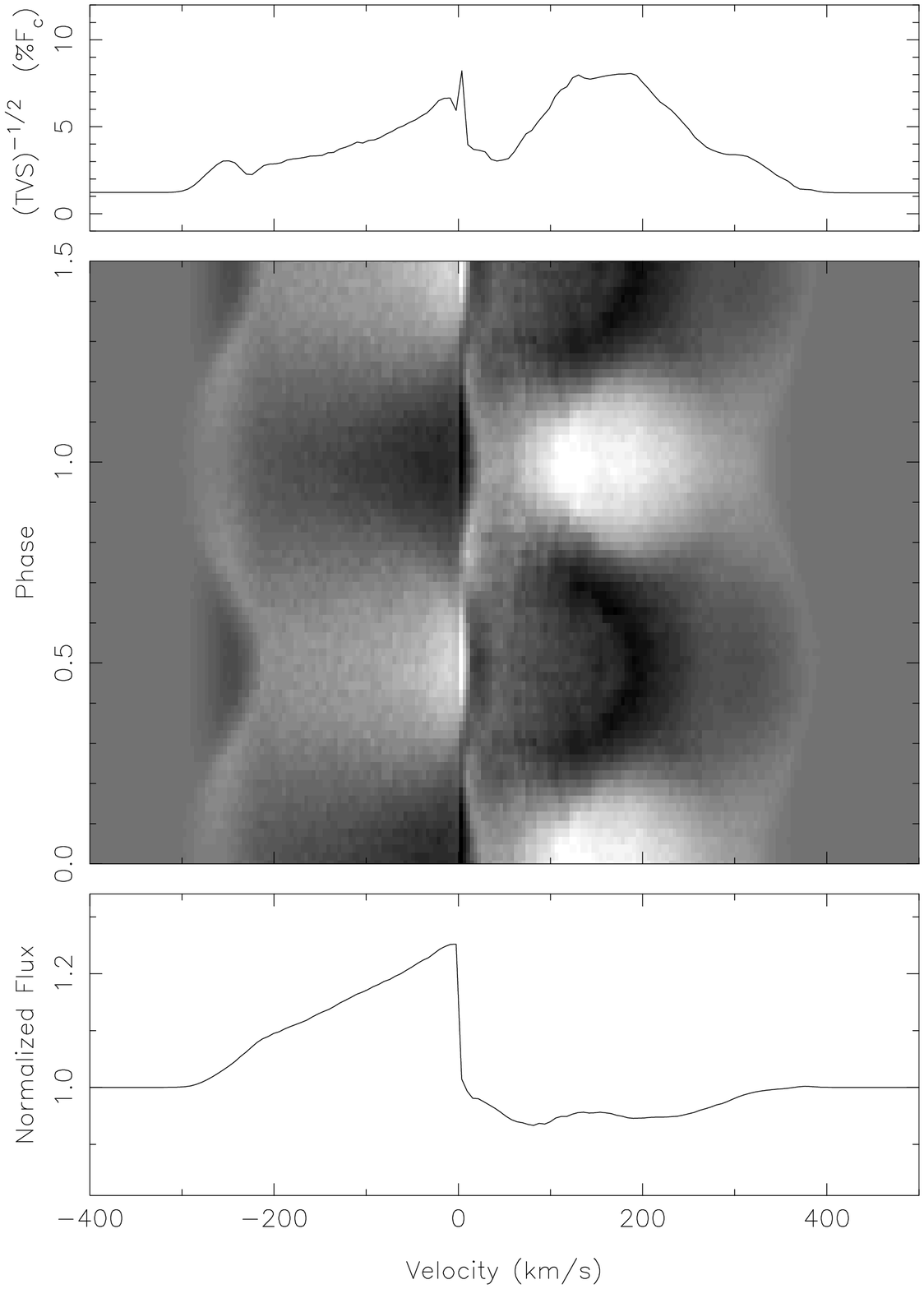}&
\includegraphics[%
  clip,
  scale=0.32]{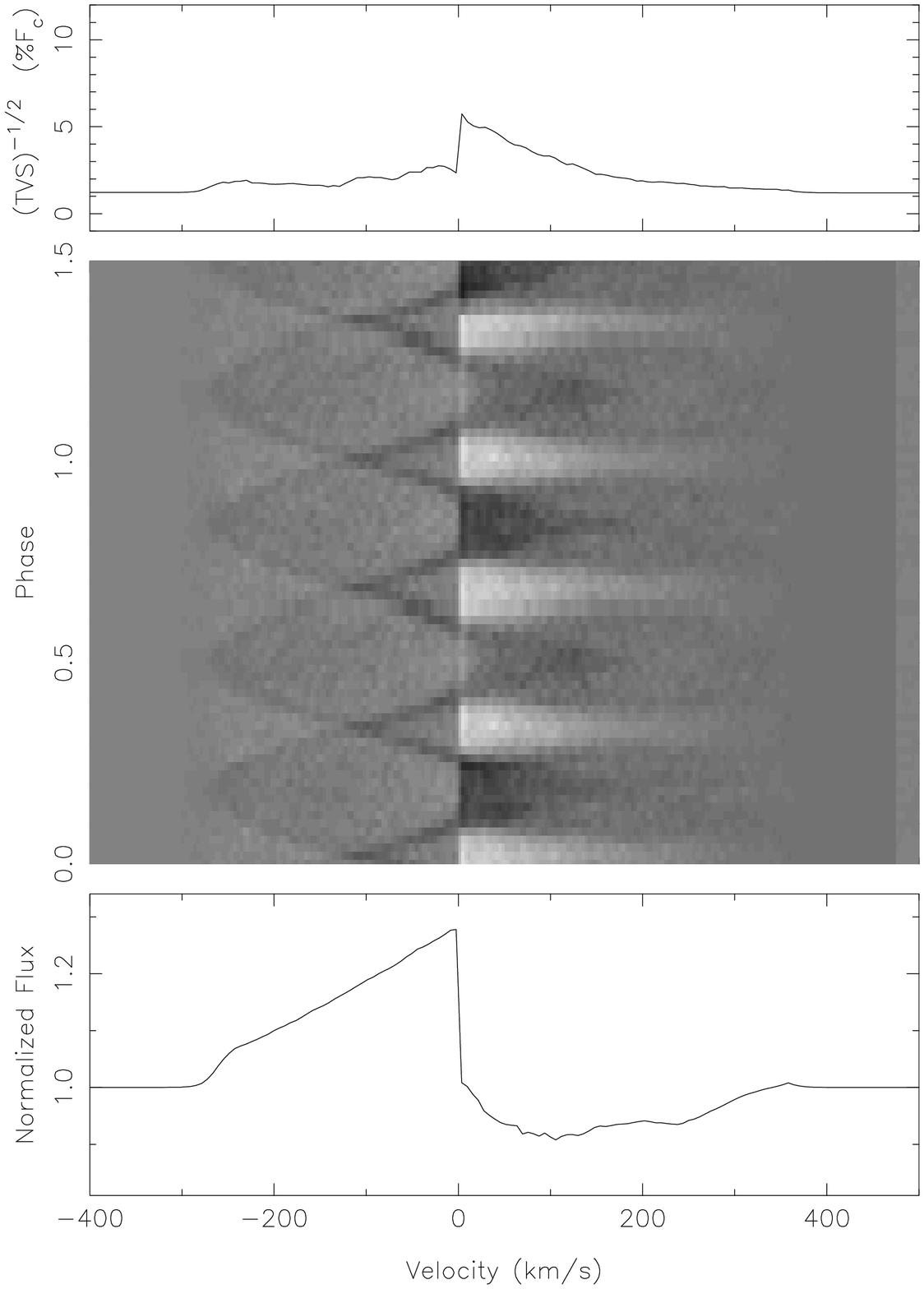}&
\includegraphics[%
  clip,
  scale=0.32]{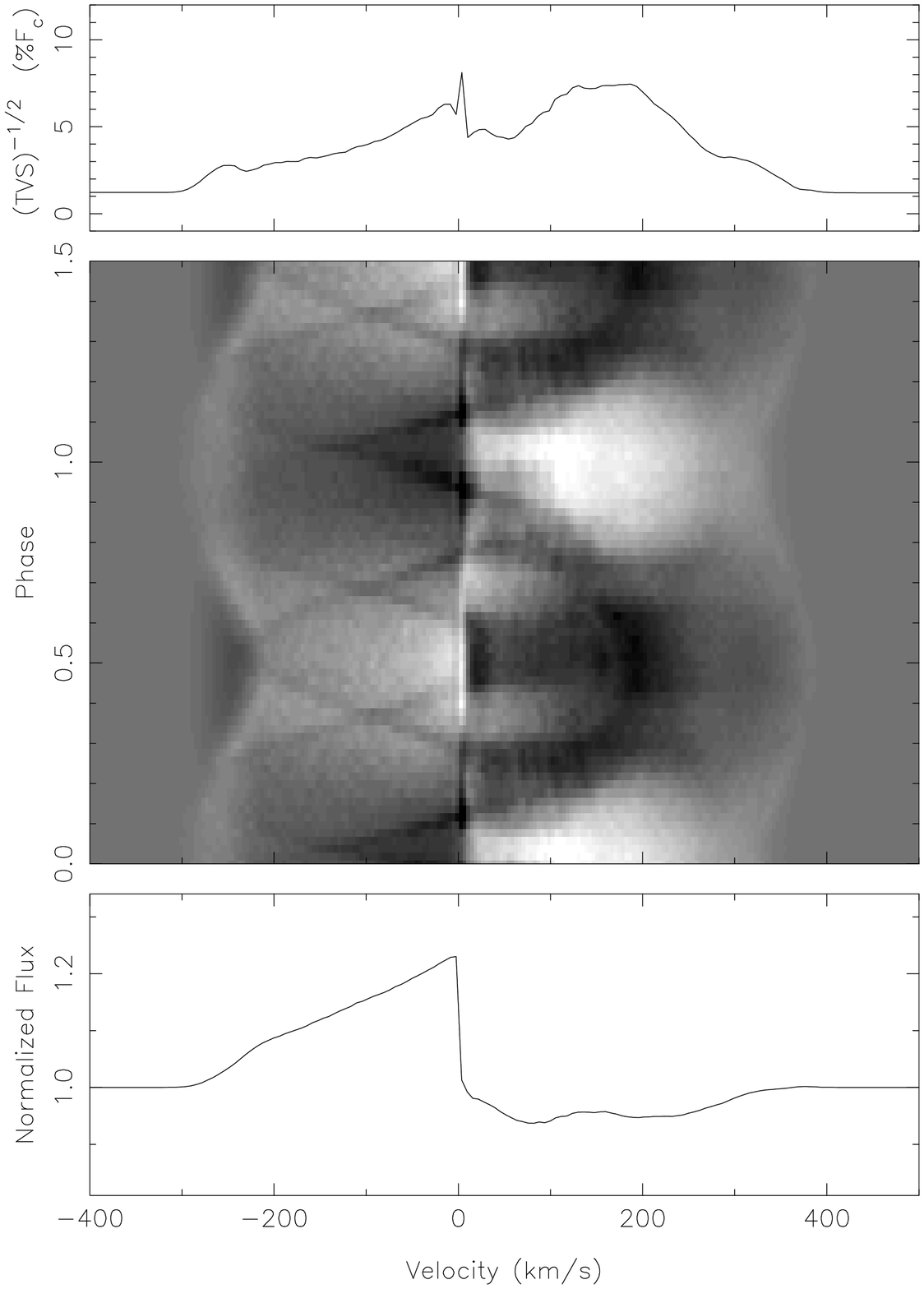}\tabularnewline
\end{tabular}

\end{center}

\caption{The summary of the Pa$\beta$ spectra computed for Model~A (left),
  Model~B (centre) and Model~C (right). For each model, spectra were
  computed at 50 different rotational phases.
  In the bottom panels, the mean spectra of all rotational
  phases are shown.   In the middle panels,
  the quotient spectra (divided by the mean spectrum) are shown as
   greyscale images with increasing rotational phases in vertical direction. 
  The greyscale image is  scaled from $1.1$ (white) to $0.9$ (black). 
  The temporal deviation spectra $(\mathrm{TVS})^{-1/2}$ are shown in
  the top panels. When computing the temporal deviation spectra,
  the signal-to-noise ratio of 90 (in continuum) was used to match that
  of the observation (Section~\ref{sec:Observations}). See
  Table~\ref{tab:model-parameters} for the model parameters adopted.}

\label{fig:model_spectra}

\end{figure*}

We have computed the Pa$\beta$ spectra at $50$
different rotational phases,  and placed the results in 
Fig.~\ref{fig:model_spectra}. The figure shows the mean spectra of a
whole phase, the deviations from the mean spectra (quotient spectra)
as a function of phase in the greyscale image, 
and the temporal deviation spectra $(\mathrm{TVS})^{-1/2}$ from the models. 

The mean spectra of the models broadly reproduce the overall shape
and relative intensity of the observations (Fig.~\ref{fig:threenights}), although the 
model line profiles  are  rather more triangular. Little
difference is seen in the shape of the mean spectra from three models.
The models overestimate the levels of the blue-wing (between $\sim
-200$ and $\sim -100\,\kmps$) compared to the observation. The
discrepancy could be caused by e.g.~the wrong inclination angle, the
wrong geometry of the magnetic field line used in the model, and the
fact that the model does not include the rotational velocity component
in the accretion streams. If a lower (e.g. $60^\circ$) inclination
angle is used, the profile will become less triangular. Although a
lower inclination model will give us a better fit to the observed mean
profile, the variability in the red absorption trough will be much
smaller than that of the observation, and the variability in the blue
side will be much larger than that of the observation. The mean
spectra from Models~A and C show the subpeak in the red absorption
trough similar to the one in seen in the observation; however the
positions of the subpeak in both models are $\sim 100~\kmps$ smaller
than that of the observed spectra.

In Model~A, the magnetic axis is perpendicular to the observer when
$\mathrm{phase}=0$. Around this phase, the subpeak in the red
absorption trough become most prominent (seen as brighter pixels in
the greyscale image). The subpeak arises from a simple geometrical
effect: at high inclination the observer's line-of-sight (LOS) towards
the hotspots passes through both low velocity material near the disc,
and high velocity material near the stellar surface, leading to two
absorption components on the red side of the line profile. The
relative position and strength of these components changes as the
magnetosphere becomes more face-on. The level of the subpeak should
become lower as the phase changes from 0 to 0.5 because of this
geometrical/projected velocity effect. The greyscale image of Model~A
shows this effect. Overall line variability seen in the red absorption
trough is well reproduced by Models~A and C. Since this variability is
primarily a geometrically effect, it is quite insensitive to the
adopted temperature structure, and we find qualitatively similar
behaviour for models with $6000\,\mathrm{K} < T_{\mathrm{max}} <
10000\,\mathrm{K}$ (although naturally the models differ in detail).

The greyscale image of Model~B show the periodic variation (with $1/3$ of a
phase) of the flux levels in the red absorption. The flux level in the
red absorption increases as a longitudinal gap approaches the line of
sight of the observer since line photons suffer less absorption. Finally, the
greyscale of Model~C shows a complicated variation pattern, but it is
easy to see that the pattern is essentially a combination of the variability
patterns from Models~A and B. 
   
The temporal deviation spectra of Models~A and C reproduce
approximately the same range of the line variability
($-400\,\kmps$,~$\,500\,\kmps$) as seen in the observations 
(Fig.~\ref{fig:threenights}). Model~B does not fit the  $(\mathrm{TVS})^{-1/2}$ of
the absorption at all.  In the $(\mathrm{TVS})^{-1/2}$ of Models~A and
C, rather large discrepancies are seen in the amount of the
variability around the line centre (slightly on the blue side). 
While the peak of the variability in the red absorption occurs at
$\sim 150\,\kmps$ for Model~A and C, it occurs at $\sim 275\,\kmps$ in
the observation.

A possible way to shift the peak of the $(\mathrm{TVS})^{-1/2}$ in the
red wing (equivalently the position of the subpeak in the mean
spectra) to a higher velocity bin is to have a larger magnetospheric
radius ($R_{\mathrm{m}}$), in which the accretion gas has higher
infalling velocity. Although this means the magnetic accretion streams
have to extend beyond the co-rotation radius of the star, this would
be still consistent with the magneto-hydrodynamic model of
\citet{romanova:2002} who found a torque-less accretion is possible
when $R_{\mathrm{m}}\approx 1.5 R_{\mathrm{c}}$ where $R_{\mathrm{c}}$
is the co-rotation radius of a star.

\subsection{Model auto-correlation maps}

\label{sub:auto-correlation-map}

\begin{figure}
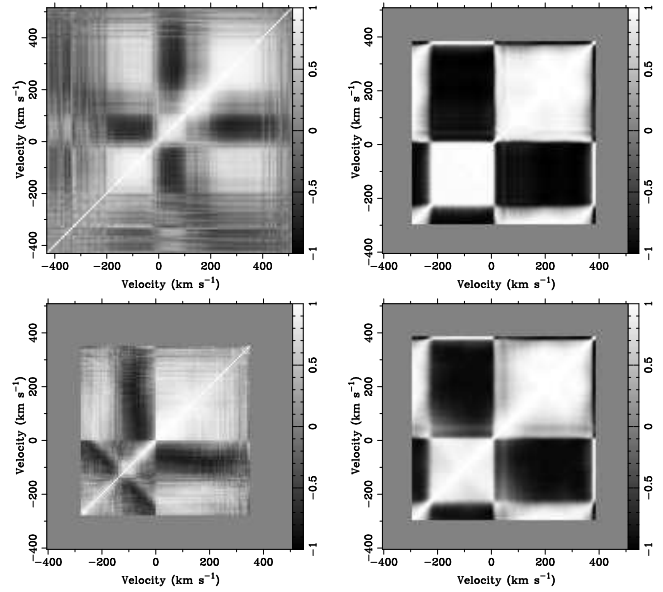



\begin{tabular}{cc}
\includegraphics[%
  scale=0.22]{fig11a.eps}&
\includegraphics[%
  clip,
  scale=0.22]{fig11b.eps}\tabularnewline
\includegraphics[%
  clip,
  scale=0.22]{fig11c.eps}&
\includegraphics[%
  clip,
  scale=0.22]{fig11d.eps}\tabularnewline
\end{tabular}


\caption{Auto-correlation maps of Pa$\beta$ from the observation
  (upper--left) and from the radiative transfer models: Models A
  (upper--right), B (lower--left) and C (lower--right). The velocity
  bins of the continuum are excluded from the model auto-correlation
  maps, and their values are set to 0 (seen as gray outer frames). No
  anti-correlation between the velocity ranges ($0\,\kmps$,\,$100\,\kmps$) and
  ($200\,\kmps$,\,$400\,\kmps$) is seen in the models unlike the
  observation. The position of the two dark stripes (anti-correlation
  bands) crossing perpendicular to each other is on the blue side
  ($-125\,\kmps$ and $-125\,\kmps$) for Models~A and C, but it
  is on the red side ($100\,\kmps$ and $100\,\kmps$), in the
  observation. } 

\label{fig:model_xcorr}

\end{figure}

The auto-correlation
maps of Pa$\beta$ line (with no time lag) were computed using the
spectra of Models~A, B and C, and the results were placed in Fig.~\ref{fig:model_xcorr}
along with the auto-correlation map shown in Fig.~\ref{fig:auto-correlation}
for a comparison. 

In the map for Model~A,  the changes in the flux levels in the
velocity range ($-200\,\kmps$,~$0\,\kmps$) and
($0\,\kmps$,~$400\,\kmps$) are strongly correlated with themselves while they
are anti-correlated with each other. In other words, as the flux level
in ($-200\,\kmps$,~$0\,\kmps$) increases the flux level in
($0\,\kmps$,~$400\,\kmps$) decreases, and vice versa. 

The blocks of correlation and anti-correlation (seen as black and
white squares in the map of Model~A) are less pronounced for
Model~B. The third quadrant block
($-300\,\kmps$,~$0\,\kmps$)$\times$($-300\,\kmps$,~$0\,\kmps$)
of Model~B is more complicated than that of Model~A because the
accretion streams located on the far (back) side of the star (as seen
by the observer) will be seen through the gap as it approaches the
near (front) side of the star.  The map for Model~C is essentially
the same as that of Model~A since the overall variability caused by the
presence of the gaps is relatively small compared to that by the
precession of the magnetic axis.   

The main differences between the auto-correlation map of the
observation and the models are: 1.~the
absence of the anti-correlation between the velocity ranges ($0\,\kmps$,~$100\,\kmps$)
and ($200\,\kmps$,~$400\,\kmps$) in the models and \textcolor{black}{2.~For Models~A and
C, the two dark stripes (anti-correlation bands) crossing perpendicular
to each other are about two times wider than that of the observation,
and they appear as two large dark blocks and two smaller blocks.}
The total phase coverage of the observation is approximately $1/3$ of
the whole phase; therefore, this may also contribute to the
difference seen in the maps from the observation and the models.

\subsection{Motion of the subpeak in the red absorption}

\label{sub:model-subpeak}

\begin{figure}

\begin{center}

\includegraphics[%
  clip,
  scale=0.45]{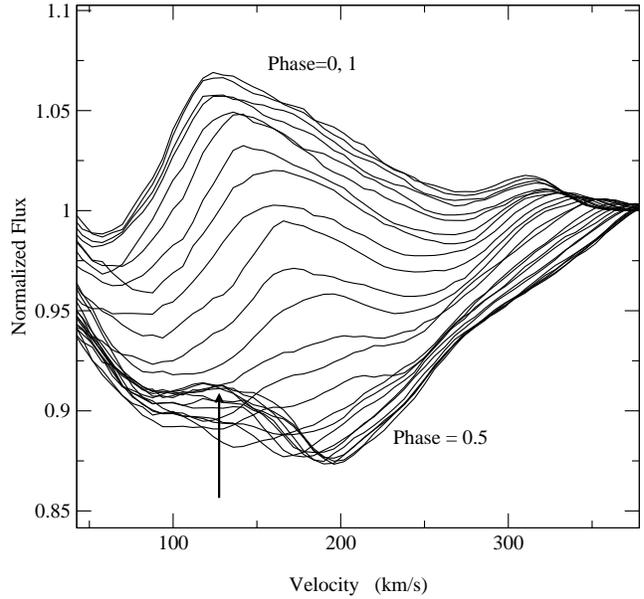}

\end{center}

\caption{The red wings ($V>0\,\kmps$) of the Pa$\beta$ spectra from 
Model~A at 25 different rotational phases (0--1). The
velocity position of the subpeak moves from $V\approx
120\,\kmps$ (at $\mathrm{phase}=0$) to $V\approx 200\,\kmps$ (at
$\mathrm{phase}=\,\sim0.25$), and then to  $V\approx 130\,\kmps$ (at
$\mathrm{phase}=0.5$). The motion is symmetric about
$\mathrm{phase}=0.5$. As the phase 
approaches $0.25$, the main subpeak feature becomes less pronounced,
and it smoothly shifts to a smaller subpeak (indicated by an arrow)
seen at $V\approx 130\,\kmps$ when the phase is around 0.5. Also see
Fig.~\ref{fig:subpeak_pos_model_A}.} 

\label{fig:subpeak_model}

\end{figure}

In Section~\ref{sub:motion-of-subpeak}, we found that the motion of the
subpeak in the red absorption trough was possibly related with the 
precession of magnetic axis around the rotational axis.  In
Fig.~\ref{fig:subpeak_model}, a subset (25 out of 50) of Pa$\beta$ 
line spectra from Model~A is shown.  Only the red side of the line profiles
are shown to emphasise the motion of the subpeak. At
$\mathrm{phase}=0$, the inclination angle of the magnetic axis with
respect to an observer is $90^{\circ}$, then it gradually decreases to
$70^{\circ}$ as the phase reaches to $0.5$. The inclination of the 
magnetic axis increases again to $90^{\circ}$ as the phase reaches to $1$.

The velocity position of the subpeak moves from $V\approx
120\,\kmps$ (at $\mathrm{phase}=0$) to $V\approx 200\,\kmps$ (at
$\mathrm{phase}=\,\sim0.25$), and then to  $V\approx 130\,\kmps$ (at
$\mathrm{phase}=0.5$). The motion is symmetric about
$\mathrm{phase}=0.5$. As the phase approaches $0.25$, the main subpeak
feature becomes less pronounced, and it smoothly shifts to a smaller
subpeak seen at $V\approx 130\,\kmps$ (indicated by an arrow in the
figure) as the phase approaches 0.5.  

\begin{figure}

\begin{center}

\includegraphics[%
  clip,
  scale=0.45]{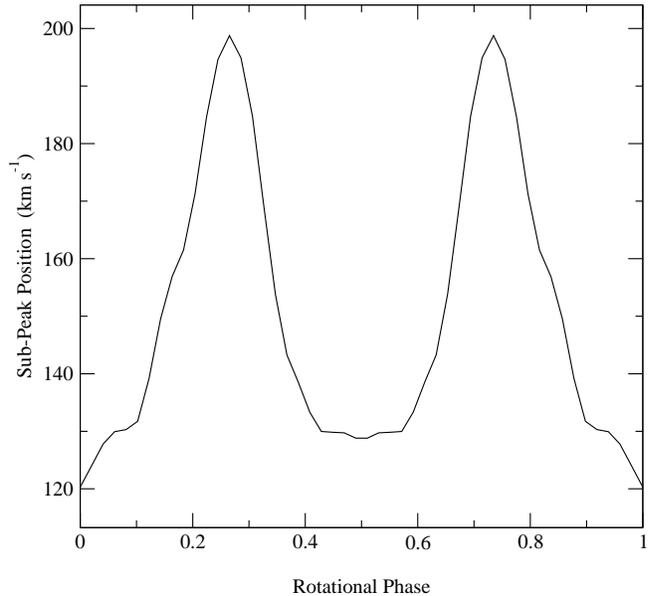}

\end{center}

\caption{The positions of the subpeak in the red wing of the Pa$\beta$
time-series spectra (Model~A) shown in Fig.\ref{fig:subpeak_model} are
plotted as a function of rotational phase (0--1).  
The figure clearly shows the presence of the second harmonic
components as seen in the observation (Fig.~\ref{fig:motion-of-subpeak}).  
At phase around  $0.25$ and $0.75$, the velocity reaches the maximum
value ($V\approx 200\,\kmps$) which is about 1.4 times smaller than
the maximum velocity seen in the observation ($V\approx
280\,\kmps$). At $\mathrm{phase}=0$ and $1$,  the velocity reaches the
minimum value ($V\approx 120\,\kmps$) which is comparable to that of
the observation ($V\approx 140\,\kmps$). }

\label{fig:subpeak_pos_model_A}

\end{figure}

Fig.~\ref{fig:subpeak_pos_model_A} shows the positions of the subpeak,
presented in Fig.\ref{fig:subpeak_model}, as a function of rotational phase.   
It clearly shows the presence of the second harmonic
components as seen in the observation (Fig.~\ref{fig:motion-of-subpeak}).  
At phase around  $0.25$ and $0.75$, the velocity reaches the maximum
value ($V\approx 200\,\kmps$) which is about 1.4 times smaller than
the maximum velocity seen in the observation ($V\approx
280\,\kmps$). At $\mathrm{phase}=0$ and $1$,  the velocity reaches the
minimum value ($V\approx 120\,\kmps$) which is comparable to that of
the observation ($V\approx 140\,\kmps$).

In the time-series spectra of Model~B, no clear subpeak is seen in the
red absorption trough therefore it is not shown here. Model~C shows a
similar variability pattern of the subpeak as in Model~A. The gaps in
the magnetosphere in Model~B and C, are unlikely the causes of the
subpeak motion.  From this analysis, we found that the Models~A and C
reproduce qualitatively similar variability in the red absorption as
seen in the observation. 

\subsection{Line equivalent width variability}

\label{sub:model-EW}
\begin{figure}

\begin{center}

\includegraphics[%
  clip,
  scale=0.48]{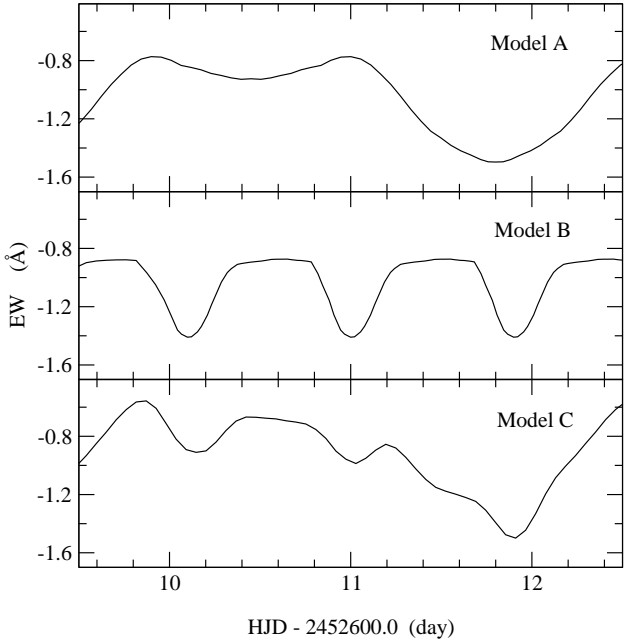}

\end{center}

\caption{ The equivalent
  widths (EWs) of Pa$\beta$ as a function of time from
  Model~A (top), Model~B (middle), and Model~C (bottom).
  The model phase was multiplied by the period of SU~Aur ($2.7\,\mathrm{d}$;
  \citealt{unruh:2004}), and then translated by $+9.1$~d to match
  the peaks of the observed EW.  Models~A and C reproduced the
  overall characteristics of the EW variability seen in the
  observation (see Fig.~\ref{fig:EW_flux_var}), but Model~B does not.}  

\label{fig:EW_model}

\end{figure}

Based on the model spectra computed in
section~\ref{sub:model-spectra}, the EW of Pa$\beta$ were computed
using the velocity bins from $-500~\kmps$ to $500~\kmps$, and
plotted as a function of time (Fig.~\ref{fig:EW_model}). 
The model phase was multiplied by the period of SU~Aur ($2.7\,\mathrm{d}$;
\citealt{unruh:2004}), and translated by $+9.1$\,d to match
approximately the positions of the peaks in the observed EW
variability curve. 

According to Fig.~\ref{fig:EW_model}, the ranges of the EW variation
predicted by Models~A, B and C are about $0.7$, $0.5$ and $0.9$\,\AA\,
respectively. On the other hand, the observation
(Fig.~\ref{fig:EW_flux_var}) shows the ranges of $0.7$, $0.5$ and
$1.2$\,\AA\, for the first, second and third nights.  The amplitudes of EW
changes predicated by the models are within the range of the observational
uncertainty ($\sim 0.3$\,\AA). A direct comparison between the 
absolute values of EW would be misleading since our model is not
a formal fit to the data. Nonetheless, we note that our model mean EW
($-1.0$\,\AA) is comparable to the mean EW of the observation
($-0.7$\,\AA), although of course the observations do not fully
sample a complete rotational period.

The overall characteristics of the EW variability seen in the
observation is well reproduced by Models~A and 
C, but Model~C with three gaps is preferred because it shows both $P/2$
and $P/3$ variation found in the observation
(section~\ref{sub:Line-equivalent-width}).  Model~B is clearly
inconsistent with the data. 
A better match of the observation with Model~C can be achieved by
carefully choosing the longitudinal position of the gaps with respect
to the tilt of the magnetic axis. The amplitude of the EW variability can
be increased by using wider gaps in the magnetosphere, and by
increasing the tilt angle of the magnetic axis. 

Interestingly, the recent light curve calculation of
\citet{romanova:2004}, who considered the flux variation caused by the
hot spots on the surface of the rotating stars in their
magneto-hydrodynamic simulations, showed the presence of second
harmonic (two peaks in one rotational period) when the inclination
angle ($i>75^{\circ}$).  This is consistent with the EW curve of our
Model~A. 

\section{Conclusions}

\label{sec:Conclusions}

We have presented the 503 time-series echelle spectra of Pa$\beta$
line of SU Aur (Fig.~\ref{fig:threenights}). From the temporal
variance spectra analysis (Section~\ref{sub:Temporal-variance-spectra}),
we found that the line displayed relatively strong variability
\textcolor{black}{ within the velocity range ($100\,\mathrm{km\,
s^{-1}},\,420\,\mathrm{km\, s^{-1}}$) in
the red absorption trough,
and weak variability within the velocity range
($-200\,\mathrm{km\,s^{-1}},\, 0\,\mathrm{km\, s^{-1}}$) in the blue
wing} while the flux level at the line centre remained almost constant.
A hint of blue-shifted absorption at about $-150$\,$\mathrm{km\, s^{-1}}$
was found on the second night. Interestingly, a similar blue-shifted
absorption feature at $-150\,\kmps$ are seen in H$\alpha$ and
H$\beta$ profiles, which modulates with the rotational period of the
star (\citealt{giampapa:1993}; \citealt{johns:1995}). 

We found the presence of a subpeak in the red absorption trough in 
each spectra (Fig.~\ref{fig:eachnight}), and its position  changes in
time (Fig.~\ref{fig:motion-of-subpeak}).  The variability seen in the
red-wing seems to be caused mainly by the motion of this subpeak. 
The position of the peak was fitted well with the Fourier series
summed up to the second harmonic term ($n_{\mathrm{max}}=2$ in
equation~\ref{eq:fourier_series}), indicating that the variability is
associated not only with the rotational period but also a half of the
rotational period.  This is consistent with the tilted-axis
magnetospheric accretion models (e.g.\,\citealt{shu:1994};
\citealt{johns:1995}). To exclude the possibility that the large
amplitude of the change in the subpeak position seen on the second 
night was caused by a single episodic event, the object must be
observed for a few rotational periods.    

The auto-correlation map (Fig.~\ref{fig:auto-correlation})
showed the profile variability in the velocity range
($-200\,\mathrm{km\, s^{-1}},0\,\mathrm{km\, s^{-1}}$) 
weakly correlates with that for ($200\,\mathrm{km\,
s^{-1}},400\,\mathrm{km\, s^{-1}}$). The variability in those two
velocity ranges weakly anti-correlates with the variability in
($0\,\mathrm{km\, s^{-1}},100\,\mathrm{km\, s^{-1}}$). The map is very
similar to that seen in the H$\beta$ auto-correlation function shown
by \citet{oliveira:2000}.

The modulations of the line equivalent
widths (EW) (Fig.~\ref{fig:EW_flux_var}) are fitted well 
with the Fourier series (equation~\ref{eq:fourier_series}) summed up
to the third harmonic terms ($n_{\mathrm{max}}=3$ in
equation~\ref{eq:fourier_series}),  indicating that the modulations
might be associated not only with the rotational period of SU~Aur and a
half of the period, but also a third of the period. 
\textcolor{black}{
Because our data
sampling period and the $P/3$ component of the Fourier analysis are very
similar to each other, it is possible that the detection of the $P/3$
contribution may be spurious.  A definitive observation requires a
multi-site campaign to continuously monitor the object for at east two
rotational periods.
}

From the comparison of the observation with the radiative transfer
models (Models~A, B and C in Section~\ref{sec:Models}), we have learnt
the followings: 

1. The inclination angle (with respect to the magnetic axis) must be
rather high \textcolor{black}{($\sim 85^{\circ}$)} to have the subpeak
(in the red wing) in emission as seen in the mean observed profile
from the third night (see Figs.~\ref{fig:eachnight} and
\ref{fig:inclination-effect}).  
\textcolor{black}{
If a more realistic accretion disc
was used in our model, instead of the geometrically thin and opaque
disc, this angle would most likely change.
}
At the lower inclination angels ($i<\sim70^{\circ}$), the models will
overestimate the variability in the blue wing when compared to the
observation.  Since the profile shape highly depends on the geometry
of the magnetosphere, these conclusions would change if
the geometry other than the one in \citet{hartmann:1994} is used.

2. The tilted magnetic axis models (Models~A and C) reproduce the
temporal deviation spectra $(\mathrm{TVS})^{-1/2}$ which have similar
ranges and shapes as  
seen in the observation (Fig.~\ref{fig:model_spectra}).  On the other
hand, the mean spectra of the models (A, B and C) are more asymmetric
and triangular than the observed profile. The discrepancies may arise from
several simplifying assumptions in the model, in particular the adoption
of  Sobolev line transfer (which is a poor approximation at the line
centre) and the neglection of turbulent broadening.

3. The motion of the subpeak in the red absorption trough seen in the
observation (Figs.~\ref{fig:eachnight} and \ref{fig:motion-of-subpeak})
was qualitatively reproduced by the tilted magnetic axis models
(Models~A and C). See Fig.~\ref{fig:subpeak_model}.  The models
reproduced the motion of the subpeak with the second harmonic
components (associated with $P/2$ where $P$ is the
rotational period), as seen in the observation.

4. The tilted magnetic axis models with the gaps in the
magnetosphere (Models~C) predicts the line equivalent width (EW)
variability patterns very similar to that seen in the observation
(Figs.~\ref{fig:EW_flux_var} and \ref{fig:EW_model}). The model
shows the $P/2$ and $P/3$ variability of the EWs as seen in the
observation.

The models presented here represent the next step in quantifying the
magnetospheric geometry of T~Tauri stars, moving from purely
qualitative descriptions of the circumstellar environment to a more
quantitative approach. The complex nature of the
problem, with inflowing and outflowing material, non-axisymmetric (and
possibly non-stationary) accretion, render a formal fit to the data
extremely difficult. Further progress will only be possible
with an extensive, coordinated observational programme, spanning a
variety of techniques and wavebands. Zeeman Doppler Imaging (ZDI;
\citealt{semel:1993}) of the surface magnetic field topology would provide
strong constraints on the magnetospheric geometry, particularly when
combined with photometric time-series. Simultaneous H$\alpha$ and Pa$\beta$
spectroscopy would also be extremely useful, since H$\alpha$ is
appears to be a good tracer of outflow emission, while we have shown
that Pa$\beta$ probes the accretion streams. Time-series H$\alpha$
linear spectropolarimetry may also provide useful new insights
(\citealt{vink:2003}; \citealt{vink:2004}).

\section*{Acknowledgements}

\textcolor{black}{We would like to thank the anonymous referee for constructive
comments and suggestions for improving the clarity of the manuscript.}
We also thank the staff of the JAC, and Paul Hirst in 
particular, for their help in obtaining the observations. RK thanks
Stuart Littlefair for help with data reduction. RK is supported by
PPARC standard grand PPA/G/S/2001/00081.

%
%
%
%
%
%


\label{lastpage}
\end{document}